\documentclass[prd,aps,12pt]{revtex4}
\usepackage{epsfig}
\def\be{\begin{equation}}
\def\ee{\end{equation}}
\def\ba{\begin{eqnarray}}
\def\ea{\end{eqnarray}}
\def\half{{1 \over 2}}
\def\Tr{{\rm Tr}}
\def\hD{\hat{D}}
\def\g5{\gamma_5}
\def\oper{{\cal O}}

\def\sp{{\cal S}}
\def\act{{\cal A}}
\begin{document}
\title{Brane-worlds and theta-vacua}
\author{S. Khlebnikov$^1$ and M. Shaposhnikov$^2$}
\affiliation{$^1$Department of Physics, Purdue University, West Lafayette,
IN 47907, USA \\
$^2$Ecole Polytechnique F\'ed\'erale de Lausanne,
Institute of Theoretical Physics,
SB ITP LPPC BSP 720,
CH-1015 Lausanne,
Switzerland}
\begin{abstract}
Reductions from odd to even dimensionalities ($5\to 4$ or $3\to 2$),
for which the effective low-energy theory contains chiral fermions,
present us with a mismatch between ultraviolet and infrared
anomalies. This applies to both local (gauge) and global currents;
here we consider the latter case. We show that the mismatch can be
explained by taking into account a change in the spectral asymmetry
of the massive modes---an odd-dimensional analog of the phenomenon
described by the Atiyah-Patodi-Singer theorem in even
dimensionalities. The result has phenomenological implications: we
present a scenario in which a QCD-like $\theta$-angle relaxes to zero 
on a certain (possibly, cosmological) timescale, despite the absence of 
any light axion-like particle.
\end{abstract}
\maketitle
\section{Introduction}
The nontrivial vacuum structure of non-Abelian gauge theories
\cite{Belavin:1975fg,'tHooft:1976up,'tHooft:1976fv,Callan:1976je,Jackiw:1976pf}
plays an important role in particle theory. It underlies
baryon-number non-conservation in electroweak theory and the
existence of $\theta$-vacua in QCD. These $\theta$-vacua present a
problem, since $\theta \neq 0$ leads to CP violation in strong
interactions, which is severely constrained by experiment. Because
the vacuum structure depends on topology of the gauge fields, it is
sensitive to the dimensionality of space-time. So, one may wonder
what happens to electroweak instantons and the $\theta$-vacua in
scenarios where the number of space-time dimensions is extended
beyond the usual four, and if perhaps a solution to the strong-CP
problem can be achieved along these lines.

Topology of gauge fields is best discussed when the space is compact.
So, in what follows, we consider only space-times of the form
\be
{\rm spacetime}_d = \sp_{d-1} \times R_1 \; ,
\label{ST}
\ee
where $\sp_{d-1}$ is a compact space, and $R_1$ corresponds to time.
The case $d=5$ is a situation that can be of phenomenological
interest, but we also consider Abelian theories in $d=3$, which are
useful models.

\begin{figure}
\epsfysize=2.5cm \centerline{\epsffile{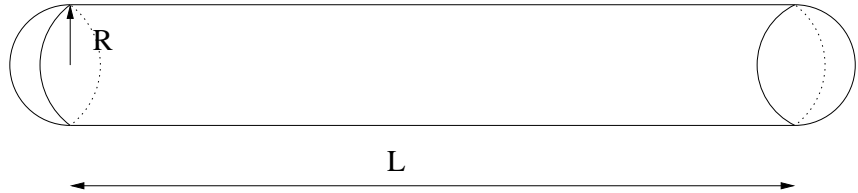}}
\caption{A sausage-like manifold leading to topologically trivial
vacuum and to the absence of $\theta$ problem.}
\label{fig:sausage}
\end{figure}

Several kinds of such higher-dimensional scenarios can be considered.
The simplest one (and, as far as we know, the first invoked in
connection with the strong-CP problem \cite{Khlebnikov:1987zg}) is
when $\sp_{d-1}$ has the topology of a 4-sphere but the geometry of a
4-dimensional sausage:
three dimensions large, and one small, see Fig. \ref{fig:sausage}.

\begin{figure}
\epsfysize=6cm \centerline{\epsffile{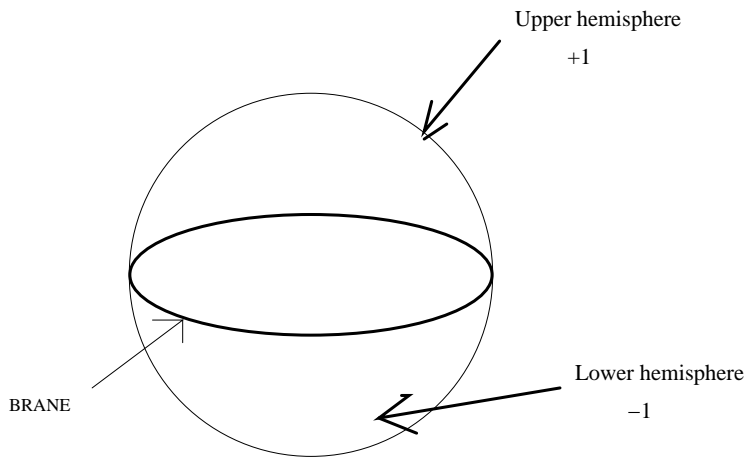}}
\caption{A manifold with the topology of a sphere and a domain
wall along the equator.} 
\label{fig:sphere}
\end{figure}

Another possibility is a brane-world: let the geometry of $\sp_{d-1}$
be more or less arbitrary---take a round 4-sphere, for example---but
suppose that we live on a domain wall along the equator see
Fig.\ref{fig:sphere} . Brane-world scenarios have been quite popular
recently, but not exactly the kind we envision here---those where
$\sp_{d-1}$ is compact. Recently, a solution to Einstein equations
with this topology was found in \cite{Gruppuso:2004db}.

\begin{figure}
\epsfysize=2.5cm \centerline{\epsffile{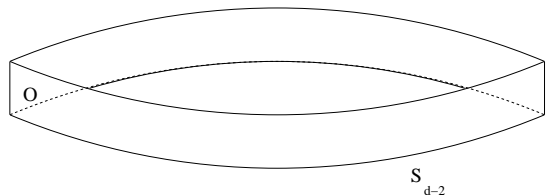}}
\caption{A space $\sp_{d-1}=O\times S_{d-2}$, where $O$ is 
an orbifold (interval).} 
\label{fig:orbifold}
\end{figure}

Finally, one can consider
$\sp_{d-1}=O \times \sp_{d-2}$, where the extra dimension is an
interval $O$, see Fig. \ref{fig:orbifold}.
In what follows, we will often call such an interval an orbifold; these
two terms will be used interchangeably.

A question closely related to topology of gauge fields is the
existence of chiral fermions and of anomalies in the corresponding
currents. Indeed, by using an anomalous chiral transformation, we can
rotate the $\theta$ angle out of the vacuum wavefunction and into the
mass matrix of the fermions. This can be sometimes a very convenient
way to represent the $\theta$ angle, since the $\theta$-dependence
can now be picked up by a calculation of the fermionic determinant.
Yet, when we try to embed this picture into a higher-dimensional
scenario, we encounter a paradox.

The lore holds that there is no anomaly in $d=3,5$ (or any other odd
dimension; in orbifold scenarios this applies in the bulk of the
orbifold, but not necessarily at the boundary
\cite{Arkani-Hamed:2001is}). This means that the chiral
transformation, which we---from our 4-dimensional
perspective---decided was anomalous, is in fact anomaly-free. Does
that mean that it can be used to safely rotate the phase of the mass
matrix to zero, without any extra terms appearing in the effective
action? If that were true, it would imply, among other things, that
any odd-dimensional theory solves the $\theta$-problem {\em
automatically}, i.e., without any reference to the theory's specific
dynamics. On the other hand, if we recall that at low energies our
odd-dimensional theory reduces to a 4-dimensional one, and so must
share its properties, this kind of automatic solution looks
exceedingly formal and suspect.

The present paper grew out of an attempt to resolve this paradox. The
solution we are going to describe reminds us of the
Atiyah-Patodi-Singer theorem \cite{Atiyah:1975jf}, in that it
emphasizes the role of high-frequency fermion modes. Although at low
energies these modes are not observable directly, changes in their
spectral asymmetry can lead to interesting low-energy consequences.
At this point, though, the similarity with the APS theorem remains
largely qualitative; in particular, they consider an even-dimensional
Dirac operator, for which there is an anomaly, while we consider an
odd-dimensional one, for which there is none. There is also some
connection between our solution and the Callan-Harvey mechanism
\cite{Callan:1984sa},  which relates the anomaly in, say, four
dimensions to a variation of a Chern-Simons term in five. However,
the Callan-Harvey mechanism reproduces a gauge anomaly, while we are
interested in a global (i.e., non-gauge) chiral transformation. The
Chern-Simons term is immune to global transformations and therefore
by itself will not do the job for us.

It is clear from the preceding that the paradox we are facing does
not depend very sensitively on whether we are considering a
non-Abelian gauge theory in five dimensions, or an Abelian theory in
three. So, in most of the paper we concentrate on the second case as
technically the simpler. With regard to the three types of
extra-dimensional models listed above, we observe that, to our
knowledge, chiral fermions have not been obtained for sausage-like
compactifications. So, in what follows, we confine ourselves to
brane-worlds and orbifolds. These two cases have many similarities
and can be treated in parallel.

For the type of questions that we address here, the global topology
of space is essential. We consider a brane-world for which the space
is a two-sphere (will be a four-sphere in the 5d version), with the
domain wall positioned along the equator.

The paper is organized as follows. In Sect. \ref{sect:ferm}, we
describe how chiral fermions appear. In Sect.
\ref{sect:current_on_orbifold}, we consider the ultraviolet
anomaly, as given by nonconservation of the current in the
odd-dimensional theory. This anomaly is zero on a sphere, while on
an orbifold it is concentrated at the endpoints
\cite{Arkani-Hamed:2001is}. We show that it can be alternatively
interpreted as a flow of charge through the endpoints.
In Sect. \ref{sect:determinant}, we compute the
infrared anomaly---the dependence of the fermion determinant on
the phase of the fermion mass. We find that it is nonzero and
coincides, in the low-energy limit, with the dependence computed
using the effective low-energy theory from the start. The mismatch
between the two anomalies is explained in Sect.
\ref{sect:spectral_asym} by considering the change in the spectral
asymmetry of massive fermion modes. Gauge-field dynamics,
responsible for existence (or non-existence) of $\theta$-vacua, is
considered in Sect. \ref{sect:gauge_field}. There, we find that
even though there is no true $\theta$-vacuum on a sphere (in
agreement with topological considerations), one can have an
effective, time-dependent $\theta$-angle. On the one hand, this
suggests a solution to the strong-CP problem; on the other, it can
have interesting cosmological consequences, if the relaxation of
$\theta_{\rm eff}$ occurs on the cosmological timescale. In Sect.
\ref{sect:disk}, we briefly discuss the case when the space is a
disk, which turns out to be similar to the case of a sphere. Sect.
\ref{sect:conclusions} is a conclusion.
\section{Chiral fermions from compact extra dimensions}
\label{sect:ferm}
\def\ctg{{\rm ctg}}
\def\ddx{\frac{d}{dx}}
\def\FF{{\cal F}}
\def\threehalfs{{3\over 2}}
For most of this section, we consider $d=3$ (corresponding to
two-dimensional``observable"  space-time). Generalization to the
realistic case $d=5$ is straightforward in the case of orbifold
and is expected to present only technical difficulties in the case
of a domain-wall on a sphere. Emergence of chiral fermions on an
orbifold is well-known in the literature, see \cite{Dixon:1985jw}
(and also \cite{Arkani-Hamed:2001is} and references therein), but
we nevertheless describe it here for completeness and to fix the
notations. Domain-wall fermions are well-known for the case when
the extra dimension is a line \cite{Rubakov:1983bb}. Here, we are
interested in the case when the higher-dimensional space is a
sphere, with the domain wall positioned along the equator. Our
analysis of this case is, as far as we know, new.

\subsection{Chiral fermion on orbifold}

Perhaps the simplest type of compactification leading to existence
of chiral fermions is related to orbifolds. Consider a 3d
space-time of the form $S_1\times O\times R_1$, where $R_1$ is
(non-compact) time, $S_1$ corresponds to large observable
dimension with size $L$ ($0<x\leq L$), and $O$ is a (short)
interval corresponding to extra dimension  ($-R/2 \leq z \leq
R/2$), $L\gg R$. The Dirac equation for the 3-dimensional
two-component fermion
\be
\Psi(t,x,z) = \left( \begin{tabular}{l} $\psi_1$\\ $\psi_2$
\end{tabular}  \right) \;
\ee
has the form
\be
i \gamma^A \partial_A \Psi + m(z)\Psi =0~, \label{dirac}
\ee
where $m(z)$ is a mass term which in general depends on the extra
coordinate $z$. We will keep $m(z)$ arbitrary as much as we can
but occasionally, for the sake of simplicity, will specialize to
$m(z) = 0$. Uppercase Latin indices scan all the coordinates,
while the Greek ones ``our" space-time.    For this subsection it
is convenient to choose the $\gamma$ matrices as follows:
$\gamma_0=\tau_1,~\gamma_1=i\tau_2$ and $\gamma_2 =i \tau_3 \equiv
i\gamma_5,~\gamma_5=\rm{diag}(1,-1)$, where $\tau_i$ are the Pauli
matrices. The signature of the metric is $(1,-1,-1)$.

The boundary condition leading to existence of a left chiral
fermion is \cite{Arkani-Hamed:2001is}
\be
(1-\gamma_5)\Psi(\pm R/2)=0~,~~\mbox{or}~~ \psi_2(\pm R/2)=0~.
\label{orbbound}
\ee
Note that since eq. (\ref{dirac}) is a system of two differential
equations of the first order, one needs exactly two boundary
conditions to specify the spectrum, as in (\ref{orbbound}). The
wave-function of the chiral fermion with momentum $k$ is simply
\be
\Psi^{(0)} \propto e^{-ik (t+x)}\left( \begin{tabular}{l} $\chi_0$\\
$0$
\end{tabular}  \right) \; ,
\ee
where $k=2 \pi l/L$ with integer $l$ because of the periodic
boundary condition $\Psi(t,0,z)=\Psi(t,L,z)$, and the zero mode is
\be
\chi_0=\frac{1}{N} \exp\left(-\int_0^z m(z')dz'\right)~,~~ N^2 =
\int_{-R/2}^{R/2}dz \exp\left(-2\int_0^z m(z')dz'\right)~.
\label{zero}
\ee

The wave functions of the Kaluza-Klein tower of massive Dirac
fermions with masses $M_n$ ($n=1,2,\dots $) are
\be
\Psi^{(n)}(z) =  \left( \begin{tabular}{l} $\chi_n(z)$\\ $\psi_n(z)$
\end{tabular}  \right) \; ,
\ee
where $\psi_n(z)$ and $\chi_n(z)$ are two sets of orthogonal
normalized functions which satisfy the equations
\ba
\nonumber \left[-\frac{d^2}{dz^2} +
m^2(z)+\frac{dm}{dz}\right]\psi_n(z)&=&M_n^2\psi_n(z)~,\\
\left[-\frac{d^2}{dz^2} +
m^2(z)-\frac{dm}{dz}\right]\chi_n(z)&=&M_n^2\chi_n(z) \label{eig}
\ea
with boundary conditions
\be
\psi_n(\pm R/2)=0~~ \mbox{and}~~
\left(\frac{d\chi_n}{dz}+m(z)\chi_n\right)\vert_{\pm R/2}=0~.
\ee
The spectrum of both operators is the same, except for the zero
mode (\ref{zero}), as they are partner Hamiltonians from the point
of view of supersymmetric quantum mechanics \cite{Cooper:1994eh}.
The relation between $\psi$ and $\chi$ is given by $\chi_n=
\frac{1}{M_n} (\partial_z -m(z))\psi_n$, $\psi_n=
\frac{1}{M_n}(-\partial_z -m(z))\chi_n$.

For the case $m(z)=0$, the wave functions have a simple form:
\be
\psi_n(z)=\sqrt{\frac{2}{R}} \sin \pi n(\frac{z}{R}
-\frac{1}{2}),~~ \chi_n(z)=\sqrt{\frac{2}{R}} \cos \pi
n(\frac{z}{R} -\frac{1}{2})~, \label{psch}
\ee
for $n=1,2,3 \dots$ and $\chi_0(z) = \frac{1}{\sqrt{R}}$. The
fermion masses are given by
\be
M_n^2=\left(\frac{\pi n}{R}\right)^2~. \label{mass}
\ee

The low-energy effective theory consists of a massless left chiral
fermion described by a one component spinor in $1+1$ dimensions.

Similar considerations apply to a theory defined in the
five-dimensional space-time $S_3\times O \times R_1$. The only
difference is that the 5d fermion has four components, while the
low-energy (4d)  massless chiral fermion now has two components.

\subsection{Chiral fermions on $S_2$}
An alternative way to obtain chiral fermions from extra dimensions
is to consider a (2+1)-dimensional theory for which the space is a
2d sphere (of unit radius). The action of a single fermionic
species is
\be
\act =\int dt\sin\theta d\theta d\phi L \; , ~~~L =
i\bar{\Psi} \hat{\partial} \Psi - \Phi(\theta) \bar{\Psi} \Psi  \; ,
\ee
where
\be
\hat{\partial} = \gamma^0\partial_0 + \gamma^1 \partial_\theta +
\gamma^2 \frac{1}{\sin\theta} (\partial_\phi + \half \gamma^1
\gamma^2 \cos\theta ) \; .
\ee
Here $\theta$ and $\phi$ are the usual polar coordinates on the
sphere, and $\Phi$ is a scalar field, whose dependence on $\theta$
is for a moment arbitrary, although later we will specify it to be
a domain wall localized on the equator (i.e., at $\theta= \pi/2$).
The field $\Psi$ is a two-component spinor: $\Psi = (\psi_1,
\psi_2)^T$. A convenient choice of $\gamma$-matrices for this
subsection is $\gamma^0 = \tau_3$, $\gamma^1 = i \tau_1$, and
$\gamma^2 = i \tau_2$. (Note that this is different from the
choice we made in the case of orbifold.)

The problem has translational symmetry with respect to time and
the azimuthal angle $\phi$, so we can take the spinor to depend on
these as $\exp[-iEt+i m\phi]$, where $m=\pm \half, \ldots$ is a
half-integer (which should not be confused with the mass $m(z)$ we
use for the orbifold theory). We then obtain the following
equations for the components:
\ba
\left[ \partial_\theta + \half \ctg \theta + {m\over \sin\theta}
\right] \psi_2
+ \Phi \psi_1 & = & E \psi_1 \; , \label{comp1} \\
\left[ \partial_\theta + \half \ctg \theta - {m\over \sin\theta}
\right] \psi_1 + \Phi \psi_2 & = & - E \psi_2 \; . \label{comp2}
\ea
These equations form the eigenvalue problem for the operator
\be
\oper = \left( \begin{tabular}{lr}
$\Phi$ & $\partial_\theta + \half \ctg \theta + {m \over \sin\theta}$ \\
$-[\partial_\theta + \half \ctg \theta - {m \over \sin\theta} ]$
& $-\Phi$
\end{tabular} \right) \; ,
\ee
whose square is
\be
\oper^2 = \left( \begin{tabular}{lr} $\Phi^2 - [\partial_\theta^2
+ \ctg\theta \partial_\theta - \frac{m^2 - m\cos\theta + {1\over
4}}{\sin^2\theta} - {1\over 4}]$ &
$-\partial_\theta\Phi$ \\
$-\partial_\theta\Phi$ & $\Phi^2 - [\partial_\theta^2 + \ctg\theta
\partial_\theta
 - \frac{m^2 + m\cos\theta + {1\over 4}}{\sin^2\theta} - {1\over 4} ]$
\end{tabular} \right) \; .
\label{O2}
\ee
We now see that the problem becomes particularly simple for $\Phi$
of the form of a step-function ($\Phi_0 > 0$):
\be
\Phi(\theta) = \left\{ \begin{tabular}{lr}
$\Phi_0 \; ,$ & $\theta < \pi/2 \; ,$ \\
$-\Phi_0 \; ,$ & $\theta > \pi/2 \; .$
\end{tabular}
\right. \label{Phi_step}
\ee
This corresponds to the limit of an infinitely thin domain wall.
In this case, the eigenvalue equation for $\oper^2$ becomes
diagonal everywhere outside the equator, while at the equator the
off-diagonal terms in $\oper^2$ produce $\delta$-function
``potentials''. We adopt this choice of the scalar-field profile
in what follows. We can then use solutions for constant fermion
mass $\Phi_0$ and match them at the equator.

Solutions for constant mass can be expressed through
hypergeometric functions, using transformations described in ref.
\cite{Abrikosov:2002jr}. In what follows, we assume that $m > 0$.
Solutions for $m < 0$ can be obtained by reflection about the
equator. Define a new coordinate variable $z = \cos^2
\frac{\theta}{2}$, and a new pair of functions $\xi(z)$ and
$\eta(z)$:
\ba
\psi_1 & = & (1-x)^{\frac{m}{2} - \frac{1}{4}} (1+x)^{\frac{m}{2}
 + \frac{1}{4}} \xi \; , \\
\psi_2 & = & (1-x)^{\frac{m}{2} + \frac{1}{4}} (1+x)^{\frac{m}{2}
- \frac{1}{4}} \eta \; ,
\ea
where $x = \cos\theta = 2 z -1$. Then, the problem reduces to the
eigenvalue problem for the operator
\be
\left( \begin{tabular}{lr} $z(1-z) \frac{d^2}{dz^2} + [m + {3\over
2} - (2m+2) z] \frac{d}{dz} -a b$ &
$-(1-z) \Phi_{,z}$ \\
$-z \Phi_{,z}$ & $z(1-z) \frac{d^2}{dz^2} + [m + {1\over 2} -
(2m+2) z] \frac{d}{dz} -a b$
\end{tabular} \right) \; ,
\ee
where
\ba
a & = & m + \half + \sqrt{E^2 - \Phi^2} \; , \\
b & = & m + \half - \sqrt{E^2 - \Phi^2} \; .
\ea
For the scalar field (\ref{Phi_step}), we can construct the
eigenfunctions $(\xi, \eta)$ at $z \leq \half$ and $z \geq \half$
from solutions to the hypergeometric equation that are regular at
the north pole  and the south pole, respectively. We obtain
\be
\xi = \left\{  \begin{tabular}{ll} $\FF(a, b, m+\threehalfs; z)$ , 
& ~~~~~~~$z \leq \half$ , \\
$\nu \FF(a, b, m + \half; 1 -z)$ , & ~~~~~~~$z \geq \half \; ,$
\end{tabular} \right. \label{xi}
\ee
and
\be
\eta =
\left\{  \begin{tabular}{ll} $-\sigma \nu \FF(a, b, m+\half; z)$ ,
& ~~~~~~~$z \leq \half$ , \\
$-\sigma \FF(a, b, m + \threehalfs; 1 -z)$ , &  ~~~~~~~$z \geq \half \;
.$ \end{tabular} \right. \label{eta}
\ee
where $\FF \equiv {}_2F_1$. From continuity,
\be
\nu = \frac{\FF(a, b, m+\threehalfs; \half)}{\FF(a, b, m + \half;
\half)} \; .
\ee
From the jump of the derivatives on the equator, we obtain
$\sigma=\pm 1$ and the eigenvalue equation
\be
\frac{\nu \FF'(a, b, m+\half; \half) + \FF'(a, b, m+\threehalfs;
\half)}{4 \Phi_0 \FF(a, b, m+\threehalfs; \half)} =\sigma = \pm
1\; , \label{spec}
\ee
which determines the allowed energies $E$.

Using the differentiation formula
\be
\FF'(a, b, m+ \half; \half)  = \frac{ab}{m+ \half} \FF(a + 1, b + 1,
m+\threehalfs; \half) \label{diff_form}
\ee
and these formulas for special values of $\FF$ \cite{Mitra}:
\begin{eqnarray*}
\FF(a, b, \half a + \half b + 1; \half) & = & 2\sqrt{\pi}
\frac{\Gamma( \half a + \half b + 1)}{b-a}
\left[ X(a,b) - X(b,a) \right] \; , \\
\FF(a, b, \half a + \half b; \half) & = & \sqrt{\pi} \Gamma( \half
a + \half b) \left[ X(a,b) + X(b,a) \right] \; ,
\end{eqnarray*}where
\be
X(a,b) = \frac{1 }{\Gamma(\half b) \Gamma(\half a + \half)} \; ,
\label{X}
\ee
and $\Gamma$ is Euler's $\Gamma$-function, we take the eigenvalue
equation (\ref{spec}) to the form
\be
b - a = \sigma \Phi_0 \left[ \frac{X(a,b)}{X(b,a)} -
\frac{X(b,a)}{X(a,b)} \right] \; . \label{spec2}
\ee
Eq. (\ref{spec2}) can be explored in considerable detail in the
limit $|a|$, $|b| \gg 1$, when we can use the expansion
\be
\frac{\Gamma(\half a + \half)}{\Gamma(\half a)} =
\sqrt{\frac{a}{2}} \left[ 1 - \frac{1}{4 a} + \frac{1}{32 a^2} +
O(a^{-3}) \right] \; . \label{expG}
\ee
In particular, this limit applies in the case of main interest to
us: $\Phi_0 \gg 1$ and $E \ll \Phi_0$, corresponding to a light
bound state on the domain wall. This state has $\sigma =1$, and
for its energy we obtain
\be
E^2 = m^2 + O(m^2 / \Phi_0^2) \; . \label{E2}
\ee
We recall that $m$ is a half-integer. In units where the radius of
the sphere is ${\cal R}$ (rather than 1), eq. (\ref{E2}) gives
$E^2 \approx m^2 / {\cal R}^2$, which is the dispersion law of a
massless fermion propagating along the equator.

The sign of $E$ can be found by returning to eqs. (\ref{comp1}), 
(\ref{comp2}). We find $E\approx -m / {\cal R}$, which corresponds to 
a left-moving, i.e., chiral fermion in (1+1) dimensions.

The transition to the effective (1+1) theory is achieved by
projecting the field $\Psi$ onto the massless mode, i.e., by
writing
\be
\Psi(z, \phi; t) = \frac{1}{\sqrt{2\pi}} \sum_m e^{im\phi} \left(
\begin{tabular}{r} $\xi_m(z)$ \\ $\eta_m(z)$ \end{tabular} \right)
A_m(t) \; , \label{proj}
\ee
where $A_m$ is the amplitude of a single-component (chiral) 2d
fermion. Note that we have indicated explicitly the dependence of
$\xi$ and $\eta$ on $m$, which was implicit before. Also, we now
assume that the basis spinor is normalized by the condition (no
sum over $m$)
\be
\int ( \xi^*_m \xi_m + \eta_m^* \eta_m )\sin\theta d\theta = 1 \;
. \label{norm_spin}
\ee
This condition makes $A_m$ canonically normalized. Note that the
fermionic mode of an opposite chirality is singular at the poles of a
sphere and is not normalizable.

In what follows, we will consider theory with two such chiral
fermions, produced by two fields $\Psi_1$ and $\Psi_2$, whose
interactions with the domain-wall field $\Phi$ have opposite
signs. If $\Psi_1=\Psi$ and is given by (\ref{proj}), then
\be
\Psi_2(z, \phi; t) = \frac{1}{\sqrt{2\pi}} \sum_m e^{im\phi}
\left( \begin{tabular}{r} $\xi_m(z)$ \\ $-\eta_m(z)$ \end{tabular}
\right) B_m(t) \; . \label{proj2}
\ee
The presence of two fields makes possible a mass term $\mu
\bar{\Psi}_1 \Psi_2$ with a complex $\mu$. Let us see what becomes
of this mass term upon the reduction to 2d. We have
\be
\int \bar{\Psi}_1 \Psi_2 \sin\theta d\theta d\phi = \sum_m
A^\dagger_m B_m \int (\xi_m^*, ~\eta_m^*) \gamma^0 \left(
\begin{tabular}{r} $\xi_m$ \\ $-\eta_m$ \end{tabular} \right)
\sin\theta d\theta \; . \label{mass_term}
\ee
Recalling that $\gamma^0= \tau_3$ and using the normalization
condition (\ref{norm_spin}), we see that the result is the
canonical mass term connecting two chiral 2d fermions.

\section{Fermion current on an orbifold}
\label{sect:current_on_orbifold} As discussed in the introduction,
one of the ingredients of the paradox that motivated the present
study is the popular assertion of the absence of anomalies in odd
dimensions ($d=\mbox{odd}$).  While for the case when the space is
a $(d-1)$-dimensional sphere we have no reason to doubt that
assertion, for the case of an orbifold the precise statement
requires some care. Namely, it is known that in that case
anomalies are absent in the bulk of the orbifold but may exist on
its boundary \cite{Arkani-Hamed:2001is}. We pause here to review
this boundary anomaly and to show that it can be interpreted as the
flow of the corresponding current through the boundary of the
orbifold.

Consider the theory of just one 3d fermion $\Psi$ with coupling
$e$ to a gauge field $A_B$, $B = 0,1,2$. The interpretation that
we are going to derive will apply also to global currents in
theories with more than one fermion species.

On the orbifold $-R/2 \leq z \leq R/2$, the single-fermion theory,
according to the calculation in ref. \cite{Arkani-Hamed:2001is},
is inconsistent as it contains a gauge anomaly concentrated at the
end points $z=\pm R/2$. It is customary to write
\be
\partial_A J^A =
\frac{e}{4\pi}\epsilon_{\mu\nu}F^{\mu\nu}\left[\delta(z-R/2)+\delta(z+R/2)\right]~,
\label{gaugean}
\ee
where $J_A = \bar{\Psi}\gamma_A\Psi$. We will show here that the
mathematical inconsistency of this theory has a simple physical
interpretation. Namely, we will demonstrate that the limit
\be
\lim_{\epsilon\rightarrow +0 }\left(J^2(R/2-\epsilon)-
J^2(-R/2+\epsilon)\right)
=\frac{e}{8\pi}\epsilon_{\mu\nu}\left[F^{\mu\nu}(-R/2)+F^{\mu\nu}(+R/2)\right]~.
\label{cur}
\ee
is non-zero in the presence of background gauge field. Note that
the values $J^2(\pm(R/2))$ {\em are equal to zero,} as this is
enforced by the boundary conditions (\ref{orbbound}). In other
words, one may either insist that the generator of the global
gauge transformation is given by
\be
Q=\int_{-R/2}^{+R/2}dz d^2x J^0
\ee
and is not conserved because of the anomaly (\ref{gaugean}), while
the flux through the endpoints of the orbifold is zero, or one may
define the charge as a limit
\be
Q=\lim_{\epsilon\rightarrow +0}
\int_{-R/2+\epsilon}^{+R/2-\epsilon}dz d^2x J^0
\ee
and relate its non-conservation to non-zero charge flux through
the endpoints.

\begin{figure}
\epsfysize=8cm \centerline{\epsffile{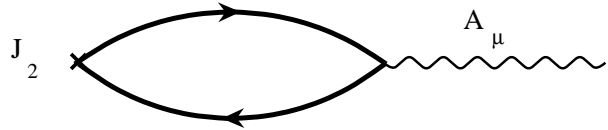}} \vspace{-3cm}
\caption{The Feynman that gives a non-zero  contribution to the
current through the endpoints of the interval.} \label{current}
\end{figure}

Let us now derive eq. (\ref{cur}). Formulas for fermionic
propagators on the orbifold are collected in Appendix A. For the
present theory, the fermionic propagator is given by $S_{11}$ with
$\mu=0$, where $S_{11}$ is defined in (\ref{propag}). The value of
$J^2$ in a background gauge field can be found from the diagram in
Fig. \ref{current} and is given by
\ba
J^2(x^\mu,z)= e \int d^2x' dz'
\epsilon^{\mu\nu}A_\mu(x',z')\times\\
\nonumber \left[
(\partial_z-m)G_D(x-x';z,z')\partial'_\nu G_D(x'-x;z',z)\right.-&& \\
\nonumber
(\partial_z+m)G_N(x-x';z,z')\partial'_\nu G_N(x'-x;z',z)+&&\\
\nonumber
\partial_\nu G_D(x-x';z,z')(\partial'_z+m)G_N(x'-x;z',z)-&&\\
\nonumber \left.
\partial_\nu G_N(x-x';z,z')(\partial'_z-m)G_D(x'-x;z',z)\right]~&&,
\ea
where $G_N$ and $G_D$ are the Green functions defined by
(\ref{green}) with $\mu=0$. With the use of (\ref{help}) this can
be simplified further to give
\ba
\label{curs} J^2(x^\mu,z)= 2e \int d^2x' dz'
\epsilon^{\mu\nu}A_\mu(x',z')\times\\
\nonumber \left[
\partial_\nu G_D(x-x';z,z')(\partial'_z+m)G_N(x'-x;z',z)\right.-&&\\
\nonumber \left.
\partial_\nu G_N(x-x';z,z')(\partial'_z-m)G_D(x'-x;z',z)\right]~&&.
\ea 
In this section we will compute the current for a background
gauge field that is independent of $z$; a more general situation---a
field slowly varying with $z$---is considered in Appendix B.

If the background field does not depend on $z$, the integration in
(\ref{curs}) over $z$ can be performed with the help of
(\ref{gdgn}) and orthogonality of the functions $\psi_n$ and
$\chi_n$. The result is
\be
J^2(x^\mu,z)=-\frac{e}{4\pi}\epsilon_{\mu\nu}F^{\mu\nu} \lambda(z)
\ee
where
\be
\lambda(z)=\sum_{n=1}^\infty\frac{\psi_n(z)\chi_n(z)}{M_n}=
\left[\frac{1}{2}\partial_z-m(z)\right]G_D(z)=
\frac{1}{2}\left(\rho(z,R/2)-\rho(-R/2,z)\right)~,
\ee
and $\rho$ is defined in eq. (\ref{defrho}).
Formally, $\lambda(\pm R/2)=0$; however, $\lim_{\epsilon\rightarrow
+0}\lambda(\pm(R/2-\epsilon)) \neq 0$. The flux of the charge through the
interval end points is
\be
\lim_{\epsilon\rightarrow +0}
(J^2(R/2-\epsilon)-J^2(-R/2+\epsilon)) =
\frac{e}{4\pi}\epsilon_{\mu\nu}F^{\mu\nu}~.
\label{const_field}
\ee
This does not depend on the function $m(z)$ and  exactly
reproduces the anomaly in eq. (\ref{gaugean}). 

Eq. (\ref{const_field}) applies when the gauge field is independent of $z$. 
In Appendix B, we consider vector potentials slowly changing with $z$ and show
that in that case the flux
depends on the value of the gauge field at the endpoints only.

\section{Computation of the fermion determinant}
\label{sect:determinant} Now, consider two species of fermions in
(2+1) dimensions coupled to an external Abelian gauge field.
Dynamics of the gauge field, and in particular the question of
existence of a $\theta$-structure, will be the subject of the
subsequent sections. Here, we consider the determinant of fermions
in an external field and, specifically, its dependence on the
phase of the fermion mass. Thus, our starting point is the
following (2+1)-dimensional Lagrangian
\be
L_\Psi = i \bar{\Psi}_1 \hD \Psi_1 + m(z) \bar{\Psi}_1 \Psi_1 + i
\bar{\Psi}_2 \hD \Psi_2 - m(z) \bar{\Psi}_2 \Psi_2 - \mu
\bar{\Psi}_1 \Psi_2 - \mu^* \bar{\Psi}_2 \Psi_1 \; , \label{L_Psi}
\ee
where $\hD= \gamma^B D_B$, $D_B$ is the covariant derivative, and
$\mu=|\mu|e^{i\theta_M}$ is a small (complex) mass. For
simplicity, we take $\mu$ to be independent of coordinates, while
the 3d mass $m(z)$ can, at the moment, depend arbitrarily on the
extra coordinate $z$. We will be interested in the dependence of
the fermion determinant on the phase of $\mu$.

We consider situations when the (2+1) theory (\ref{L_Psi}) reduces
at low energies to a (1+1)-dimensional theory. In other words,
there will be a quasi-zero fermionic mode, while all other modes
will be separated from it by a large gap. This can be arranged in
both of the extra-dimensional scenarios considered in sect.
\ref{sect:ferm}.

The resulting (1+1)-dimensional theory has a pair of chiral
fermions, forming a two-component Dirac spinor of mass $\mu$. In
fact, this theory is nothing but the fermionic sector of a massive
Schwinger model in $1+1$ dimensions, with the Lagrangian
\be
L_{2d} = \bar{\psi} i\gamma^\mu D_\mu\psi -\mu \bar{\psi}_L\psi_R
- \mu^* \bar{\psi}_R\psi_L \; .
\ee
If we start from five dimensions and a non-Abelian gauge group
(say, SU(3)) we will similarly get the massive fermions of quantum
chromodynamics.

The global current of the theory (\ref{L_Psi})
\be
J_G^A  =  \bar{\Psi}_1\gamma^A \Psi_1-\bar{\Psi}_2\gamma^A \Psi_2
\label{JA}
\ee
becomes, at low energies, the chiral current of the effective
(1+1) theory:
\be
J_5^\mu  =  \bar{\psi}\gamma^\mu \gamma_5\psi \; . \label{Jmu}
\ee
In (1+1), the chiral symmetry, in addition to being broken by the
mass $\mu$, is also broken by the anomaly, which manifests itself
in a dependence of the (1+1) determinant on $\theta_M = \arg \mu$,
a dependence that does not disappear in the limit $\mu\to 0$. A
naive expectation would be that this $\theta_M$-dependence carries
over to the full (2+1)-dimensional theory. Indeed, the masses of
the heavy modes depend only weakly on $\mu$, and therefore any
contribution they make to the determinant should be regular in
$\mu$, i.e., independent of $\theta_M$ at $\mu \to 0$.

In this section, we show that the naive reasoning is in fact
entirely correct (in particular, it is not affected by ultraviolet
divergences). Thus, even though the complete (2+1) theory has no
anomaly in the (bulk) chiral current, the low energy
manifestations of the phase $\theta_M$ are the same as in the
(1+1) theory, which has such an anomaly. On the one hand, this
resolves the paradox formulated in the Introduction, but on the
other, indicates that in brane-world scenarios the breakdown of
chiral symmetry is realized rather non-trivially. Namely,
non-conservation of the fermion number is not simply counted by
the anomaly: we are pointed towards an additional effect, having
to do with the spectral asymmetry.

\subsection{Fermion determinant on an orbifold}
To obtain an effective (1+1) theory that is free of a gauge
anomaly, from an orbifold compactification, we use different
boundary conditions for $\Psi_1$ and $\Psi_2$: for $\Psi_1$, they
are those of eq. (\ref{orbbound}) whereas for $\Psi_2$ they single
out the right-handed fermion,
\be
(1+\gamma_5)\Psi_2(\pm R/2)=0~. \label{right}
\ee
The mixing of the left-handed fermion $\psi_L$ in $\Psi_1$ and
right-handed fermion $\psi_R$ in $\Psi_2$ produces a Dirac fermion
$\psi$ with mass $\mu$. Note that the 3d masses of $\Psi_1$ and
$\Psi_2$ in (\ref{L_Psi}) are the same (up to a sign), which
allows one to use the eigenfunctions defined in eq. (\ref{eig}).

Both the complete 3d and effective 2d theories are free from gauge
anomalies. However, the global currents (\ref{JA}) and (\ref{Jmu})
are anomalous: at $\mu = 0$,
\ba
\partial_A J_G^A & = &
\frac{e}{2\pi}\epsilon_{\mu\nu}F^{\mu\nu}\left[\delta(z-R/2)
+\delta(z+R/2)\right] \; , \label{glan} \\
\partial_\mu J_5^\mu & = & \frac{e}{2\pi}\epsilon_{\mu\nu}F^{\mu\nu} \; ,
\ea
where the $\delta$-function at the boundary is defined so that its
integral over $z$ is equal to $1/2$. The covariant derivative in
this subsection is $D_B = \partial_B - ie A_B$.

As discussed in Sect. \ref{sect:current_on_orbifold}, the 3d
anomaly (\ref{glan}) is concentrated at the boundary of the
orbifold. This anomaly, which determines non-conservation of the
current, will be referred to as the ultraviolet anomaly. We now
wish to see if it matches the ``infrared'' anomaly, which comes
from the dependence of the fermionic determinant on the phase
$\theta_M$.

Consider the variation of the vacuum energy with respect to
$\theta_M$ in a slowly-varying gauge field background in three dimensions. 
It is given by the diagram in Fig. \ref{current}, which can be
immediately computed with the result
\be
\frac{\partial \Omega}{\partial \theta_M}\vert_{\theta_M=0}= e\int
d^2x dz~ \epsilon_{\mu\nu}F^{\mu\nu}(x,z)\kappa(z)~,
\label{anom_slow2}
\ee
where
\be
\kappa(z) = |\mu|^2\int
d^2x'dz'\left[G_N(x';z,z')^2-G_D(x';z,z')^2\right] \; .
\ee
This result is valid for arbitrary $m(z)\neq 0$. (Definitions of various 
Green functions are given in Appendix A.)

With the use of the mode expansion this can be rewritten as
\be
\kappa(z) =|\mu|^2 \int \frac{d^2p}{(2\pi)^2}
\sum_{m=0}\frac{\chi_m(z)^2-\psi_m(z)^2}{(p^2-M_m^2-|\mu|^2)^2}=
\frac{|\mu|^2}{4\pi}\left[\tilde G_N(0,z,z)-\tilde
G_D(0,z,z)\right]~.
\ee
The function $\kappa(z)$ has the following important property
\be
\int dz \kappa(z)= \frac{1}{4\pi}~,
\label{kappa}
\ee
which shows that for $z$-independent field strengths the $\theta_M$
dependence of the vacuum energy is given entirely by the
``ultraviolet'' anomaly. Indeed, in this case, we can pull $F_{\mu\nu}$
out of the integral over $z$ in (\ref{anom_slow2}) and use (\ref{kappa}) to
obtain
\be
\frac{\partial \Omega}{\partial \theta_M}\vert_{\theta_M=0}= \frac{e}{4\pi}
\int d^2x~  \epsilon_{\mu\nu}F^{\mu\nu} \; .
\label{anom_slow}
\ee

However, for arbitrary $z$-dependent background fields, that is no
longer true. In this case, the $\theta_M$ dependence is more
complicated. For example, for a theory with $m(z)=0$ one finds,
with the help of equations from Appendix A, that
\be
\kappa(z)= \frac{1}{4\pi} \frac{\mu}{\sinh\mu R}\cosh 2\mu z \; ,
\ee
so that
\be
\frac{\partial \Omega}{\partial
\theta_M}\vert_{\theta_M=0}=\frac{e}{8\pi}\left( \int d^2x
\epsilon_{\mu\nu}\left[F^{\mu\nu}(x,-R/2)+F^{\mu\nu}(x,R/2)\right]
-\int d^2x dz \frac{\sinh 2\mu z}{\sinh\mu R}
\epsilon_{\mu\nu}\partial_z F^{\mu\nu}(x,z)\right)~.
\label{anom_orb}
\ee
The first term is a boundary contribution that can be seen to
match the ``ultraviolet'' anomaly (\ref{glan}). However, the
second---bulk---term is new. It represents a mismatch
between the ``ultraviolet'' and ``infrared'' anomalies for the case
of orbifold. For example, if $F_{\mu\nu}$ vanishes at the endpoints,
the ``ultraviolet'' anomaly is zero, but the bulk contribution still
persists.

\subsection{Determinant of domain-wall fermions in infinite flat space}
\def\MM{{\cal M}}
\def\tr{\mbox{tr}}
\def\const{{\rm const.}}
Before we consider a domain wall on the equator of a sphere, let
us look at a simpler case that has all the relevant features---a
domain wall in flat space with an infinite extra dimension. In
other words, instead of $\sp_{d-1} = S_2$, we consider
\be
\sp_{d-1} = R_1 \times S_1 \; . \label{line}
\ee
The line $R_1$ is the extra dimension. Such a theory holds no
promise for solving the strong-CP problem, but the structure of
the fermion determinant is very similar to that on a sphere. In
fact, after we handle the case (\ref{line}), transition to a
sphere will be relatively straightforward.

In this subsection, we absorb charge $(-e)$ into the field $A_B$,
so that the covariant derivative is $D_B =\partial_B + i A_B$.
Also, the 3d mass in (\ref{L_Psi}) is assumed to be entirely due
to the coupling with the domain-wall field $\Phi$:
\be
m(z) = -\Phi(z) \; .
\ee

Fermion determinant produces the following contribution to the
effective action of the gauge field:
\be
\Delta \act = -i \Tr \ln \left( \begin{tabular}{lr}
$i \hat{D} - \Phi$ & $- \mu$ \\
$- \mu^*$ & $i \hat{D} + \Phi$ \end{tabular} \right) \equiv -i \Tr
\ln \MM \; . \label{Seff}
\ee
We are interested in the derivatives of this action with respect to
real and imaginary parts of $\mu = \mu_R + i \mu_I$ or, more
precisely, in the dependence of these derivatives on the gauge
field $A$, for example,
\be
\frac{\partial}{\partial \mu_R} \Delta \act - [\ldots]_{A=0} = i \Tr
\left\{ \MM^{-1} \left( \begin{tabular}{lr} 0 & 1 \\ 1 & 0
\end{tabular} \right) - [\ldots]_{A=0} \right\} \; . \label{der}
\ee
We use notation $- [\ldots]_{A=0}$ to denote a subtraction at zero
$A_B$.

To invert the operator $\MM$, we write
\be
\left( \begin{tabular}{lr} $i \hat{D} + \Phi$ & $ \mu$ \\
$ \mu^*$ & $i \hat{D} - \Phi$ \end{tabular} \right) \MM =
\left( \begin{tabular}{lr} $(i\hD+\Phi)(i\hD-\Phi) - |\mu|^2$ & 0 \\
0 & $(i\hD-\Phi)(i\hD+\Phi) - |\mu|^2$ \end{tabular} \right)
\label{invert}
\ee
and then compute
\ba
(i\hD+\Phi)(i\hD-\Phi) & = & -D^B D_B - \half \epsilon^{ABC}
\gamma_A F_{BC}
+ \gamma^5 \Phi' - \Phi^2 \; , \\
(i\hD-\Phi)(i\hD+\Phi) & = & -D^B D_B - \half \epsilon^{ABC}
\gamma_A F_{BC} -\gamma^5 \Phi' - \Phi^2
\ea
($\epsilon^{012} = 1$). We number coordinates in the way
consistent with Sect. \ref{sect:ferm}: the extra coordinate $z$
corresponds to $B=1$, while ``our'' coordinate $x$ to $B=2$. In
this computation we have assumed that the scalar field $\Phi$
depends only on $z$, so that $\Phi' = \partial_z \Phi$. The choice
of $\gamma$-matrices is the same as in Sect. \ref{sect:ferm}:
$\gamma^0 = \tau_3$, $\gamma^1 = i \tau_1$, and $\gamma^2 = i
\tau_2$. In addition, we have introduced $\gamma_5 = -i\gamma^1$;
this will be the $\gamma_5$ matrix of the effective
(1+1)-dimensional theory.

We can now use (\ref{invert}) to express $\MM^{-1}$ through the
inverse of the operator on the right-hand side. This operator is
diagonal in ``isospin'', i.e., the index distinguishing the two
species, $\Psi_1$ and $\Psi_2$. The isospin trace can then be
found explicitly; we continue to denote the remaining spin and
coordinate trace by $\Tr$.

We will only need the derivative (\ref{der}) to the first order in
$F_{BC}$. We find that to this order it can be written as a sum of
two pieces:
\be
\frac{\partial}{\partial \mu_R} \Delta \act - [\ldots]_{A=0} = I_1 +
I_2 \; ,
\ee
where
\be
I_1 = i \Tr \left\{ \frac{\mu^*}{-D^2 - \gamma^5 \Phi' - M^2} +
\frac{\mu}{-D^2 +  \gamma^5 \Phi' - M^2} \right\} - [\ldots]_{A=0}
\; , \label{I1}
\ee
\be
I_2 = {i\over 2} \Tr \epsilon^{ABC} \gamma_A F_{BC} \left\{
\frac{\mu^*}{(\partial^2 + \gamma^5 \Phi' + M^2)^2} +
\frac{\mu}{(\partial^2 - \gamma^5 \Phi' + M^2)^2} \right\} \; .
\label{I2}
\ee
with $M^2 = \Phi^2 + |\mu|^2$. We now consider these two pieces in
turn.

{\bf Calculation of $I_1$.} This term reflects the coupling of the
gauge field to the translational motion of fermions. As expected,
no anomaly comes from this coupling; nevertheless, for
completeness, we describe the calculation in some detail.

Define $P_B = iD_B$ and consider traces of various powers of the
operator
\be
\oper = P^2 - \g5 \Phi' - M^2 \; .
\ee
In eq. (\ref{I1}) we need $\Tr \oper^{-1} - [\ldots]_{A=0}$ (and
an analogous trace with $\Phi' \to -\Phi'$). An anomaly in $I_1$
would correspond to a non-analytic behavior in the limit $\mu \to
0$: for a slowly varying $F = F_{0x}$, we would have
\be
\Tr \oper^{-1} - [\ldots]_{A=0} \approx \frac{\const}{|\mu|^2}
\int dx dt F_{0x}(0,x,t) \label{oper_1}
\ee
(assuming that the domain wall is at $z=0$). Since traces of
higher powers of $\oper^{-1}$ can be obtained by differentiating
with respect to $|\mu|^2$, they would have similar singular
limits, for example,
\be
\Tr \oper^{-2} - [\ldots]_{A=0} \approx - \frac{\const}{|\mu|^4}
\int dx dt F_{0x}(0,x,t)~. \label{oper_2}
\ee
If we find that at least one of these traces does not have the
requisite behavior, that means that the constant in (\ref{oper_1})
is in fact zero.

To verify the presence (or, rather, the absence) of these singular
contributions, we use the ``shift'' method described in ref.
\cite{Novikov:1983gd}. In the presence of a domain wall, this
method needs to be slightly generalized. In particular, we use
only shift vectors $q$ that lie within ``our'' (1+1)-dimensional
subspace.

Consider
\be
\Tr h \ln [(P -q)^2 - \g5 \Phi' - M^2] -  [\ldots]_{A=0} = \Tr h
\ln [ \oper - 2 P q + q^2 ] -  [\ldots]_{A=0} \; , \label{shift}
\ee
where $q$ is an arbitrary constant (1+1) vector, and $h$ is an
operator that depends only on the $z$ component of $P$ and so is
immune to the shift. Expression (\ref{shift}), as well as the
above expressions (\ref{oper_1}) and (\ref{oper_2}), is assumed to
be properly regularized in the ultraviolet. For example, we can
use a set of Pauli-Villars regulators. Such a regularization will
be assumed in what follows, but it will not be indicated
explicitly. The final result will be ultraviolet-finite.

Expanding in $q$ to the second order, we obtain
\be
\ln [ \oper - 2 P q + q^2 ] =  \ln \oper +  \oper^{-1} (-2Pq +
q^2) - \half \oper^{-1} (-2Pq)  \oper^{-1} (-2Pq) + \ldots \; .
\label{exp_in_q}
\ee
The idea of the ``shift'' method \cite{Novikov:1983gd}
is that, since the regularized
trace is independent of $q$, traces of the order $q^2$ terms in
eq. (\ref{exp_in_q}) should add up to zero. Averaging over
directions of $q$, we see that this leads to
\be
\Tr h \oper^{-1} - [\ldots]_{A=0} = \Tr h \oper^{-1} P_\mu
\oper^{-1} P^\mu - [\ldots]_{A=0}\; , \label{q2_terms}
\ee
where $\mu$ takes values 0 and 2. Using the commutators
\be
[ P_\mu , \oper^{-1}] = - \oper^{-1} [P_\mu, \oper ] \oper^{-1} \;
,
\ee
and
\be
[P_\mu, \oper ] = [ P_\mu, P^2] = - i \{ F_{\mu B}, P^B \} \; ,
\ee
where the braces denote an anti-commutator, we can rewrite eq.
(\ref{q2_terms}) as
\be
[\Tr h \oper^{-1}  -  \Tr h \oper^{-2} P^\mu P_\mu] -
[\ldots]_{A=0} = i \Tr h \oper^{-2} \{ F_{\mu B}, P^B \}
\oper^{-1} P^{\mu} \; . \label{fin}
\ee
The difference of the traces on the left-hand side can be
rewritten as
\be
\Tr h \oper^{-1} - \Tr h \oper^{-2} P^\mu P_\mu = \Tr h
\oper^{-2}( -P_z^2 - \gamma_5 \Phi' - M^2) \; , \label{p_x^2}
\ee
so if we choose
\be
h = ( -P_z^2 - \gamma_5 \Phi' - M^2)^{-1} \; ,
\ee
eq. (\ref{fin}) becomes
\be
\Tr \oper^{-2} - [\ldots]_{A=0} = i \Tr h \oper^{-2} \{ F_{\mu B},
P^B \} \oper^{-1} P^{\mu} \; . \label{no_anom}
\ee
This is to be compared to the would-be anomalous behavior, eq.
(\ref{oper_2}). By inspection of the right-hand side of
(\ref{no_anom}), we find that the anomalous term is absent. We
conclude that there is no anomaly in $I_1$.

{\bf Calculation of $I_2$.} This term reflect the coupling of the
gauge field to the spin of fermions, which is the coupling that
usually leads to an anomaly. In our case, the calculation of
(\ref{I2}) in the limit of a slowly varying $F$ amounts to a study
of the spectra of two effective one-dimensional Hamiltonians: $H_1
= -\partial_z^2 + \Phi^2 - \gamma_5 \Phi'$ and $H_2 =
-\partial_z^2 + \Phi^2 + \gamma_5\Phi'$. These Hamiltonians are
supersymmetric partners, and in addition both commute with
$\gamma_5$. So, their spectra can be analyzed in some detail.
However, for our present purposes, we only need the infrared parts
of the spectra. In the presence of a domain wall of $\Phi$, $H_1$
and $H_2$ each have a zero mode, with opposite chiralities. These
are the only modes that give a singular contribution in the limit
$\mu\to 0$. Therefore, in this limit, for slowly-varying 
(in comparison with $|\mu|$) fields, we obtain
\be
I_2 \approx 2i \mu_I \int dx dt F_{0x} (0, x, t) \int
\frac{d\omega d k_x}{(2\pi)^2} \frac{1}{(\omega^2 - k_x^2 -
|\mu|^2+i\epsilon)^2 } \; ,
\ee
which is the anomaly.

Combining the above results for $I_1$ and $I_2$, we find that the
anomalous term in the effective action is
\be
(\Delta \act)_{\rm anom} = \frac{\theta_M}{2\pi} \int dx dt F_{0x}
(0, x, t) \; , \label{S_anom}
\ee
where $\theta_M = \arg \mu$. This is precisely the same anomaly
that would obtained in the effective (1+1) theory describing
chiral fermions on the wall:
\be
L_{2d} = i \bar{\psi} \gamma^\mu D_\mu \psi - \mu_R \bar{\psi}
\psi - i\mu_I \bar{\psi} \gamma_5 \psi \; .
\ee

\subsection{Determinant of domain-wall fermions on a sphere}
On a sphere, the covariant derivative is
\be
\hat{D} = \gamma^0 (\partial_0 + i A_0) + \gamma^1
(\partial_\theta + i A_\theta) + \gamma^2 \frac{1}{\sin\theta}
(\partial_\phi + \half \gamma^1 \gamma^2 \cos\theta + i A_\phi) \;
.
\ee
The relevant infrared limit now is
\be
{\cal R}^{-1} \ll |\mu| \ll \Phi_0 \; , \label{limit_sphere}
\ee
where ${\cal R} = 1$ is the radius of the sphere, and $\Phi_0$ the
magnitude of the scalar field away from the equator. Because of
the explicit dependence of $\hat{D}$ on the polar angle $\theta$,
various additional terms appear in the calculation of the
determinant. Nevertheless, in the limit (\ref{limit_sphere}), the
final answer is the natural adaptation of eq. (\ref{S_anom}):
\be
(\Delta \act)_{\rm anom} \approx \frac{\theta_M}{2\pi} \int d\phi dt
F_{0\phi} (\frac{\pi}{2}, \phi, t) \; . \label{anom_sphere}
\ee

\subsection{Limit of a thin orbifold}
The orbifold and domain-wall results are related to each other. To
see that, consider the limit when the orbifold becomes thin:
$|\mu| R \ll 1$. Restricting ourselves to the case $m(z)=0$, for which
the explicit formula (\ref{anom_orb}) was obtained, we see that in 
the limit $|\mu| R \ll 1$ we can approximate the $\sinh$
functions in (\ref{anom_orb}) by their arguments and then
integrate over $z$ by parts. The boundary terms cancel, and we
obtain
\be
\frac{\partial \Omega}{\partial
\theta_M}\vert_{\theta_M=0}=\frac{e}{4\pi R} \int d^2x dz
\epsilon_{\mu\nu}  F^{\mu\nu}(x,z) \; . \label{anom_thin}
\ee
This agrees with the effective
action (\ref{S_anom}) of the domain-wall scenario, with the role of
$F_{0x}(0, x, t)$ now being played by the average of $F_{0x}$ over
the extra dimension. Thus, in a
sense, in the thin-orbifold limit, the entire orbifold plays the
role of a domain wall.

\section{Spectral asymmetry}
\label{sect:spectral_asym} We have seen that, in all of our
examples, the $\theta$-dependence of the $d$-dimensional theory
agrees with that calculated using the low-energy
$(d-1)$-dimensional fields alone, and disagrees with what one
might expect from the anomaly equation for the $d$-dimensional
current. In other words, there is a mismatch between the
``ultraviolet'' and ``infrared'' anomalies.

This mismatch implies that the anomalous production of fermions is
not counted correctly by the $d$-dimensional anomaly. The
situation is analogous to that described by the
Atiyah-Patodi-Singer theorem for a Dirac operator in even
dimensions \cite{Atiyah:1975jf}, see also ref.
\cite{Ninomiya:1984ge}. There, the index of the Dirac operator is
not given simply by the anomaly equation, but includes an
additional term (the $\eta$-invariant) having to do with the
change in spectral asymmetry. In this section, we show that a
similar mechanism is at work in our odd-dimensional theories.

The argument is the simplest when the space is a two-sphere (the
total dimensionality of space-time is $d=3$). As seen from eq.
(\ref{anom_sphere}), in this case, the $\theta$-dependence is
activated by fluctuations that change the integral of $A_\phi$
around the equator:
\be
\int d\phi dt F_{0\phi} = \int d\phi A_\phi (t_2) - \int d\phi
A_\phi (t_1) \neq 0 \; ,
\ee
where $t_1$ and $t_2$ are some initial and final times.
For brevity, we will refer to such fluctuations as ``instantons'',
even though they do not have to be associated with tunneling and
may as well take place in real time.

Now, on a sphere, the integral of $A_\phi$ along the equator
equals the magnetic flux through the northern hemisphere:
\be
\int d\phi A_\phi = \int d\phi \int_0^{\pi/2} b \sin\theta d\theta
\; ,
\ee
where
\be
b = \frac{1}{\sin\theta} F_{\theta\phi} = \frac{1}{\sin\theta}
(\partial_\theta A_{\phi} - \partial_\phi A_\theta ) \; .
\label{b_field}
\ee
It will be convenient to visualize the transport of flux as motion
of particle-like flux quanta---vortices. Flux can be localized into
vortices, for instance, by introduction of a suitable Higgs
field.

We will be interested in scattering of vortices off the domain
wall (positioned along the equator). Consider the process when a
vortex-antivortex pair is created from vacuum in the southern
hemisphere, and then the vortex is transported across the equator
to the northern hemisphere, while the antivortex remains where it
was. This changes $\frac{1}{2\pi}\int d\phi A_\phi$ by one. The
energetics of this process does not concern us at present; it will
be the subject of the next section. Here we simply assume that the
vortices are light enough to be a part of our low-energy theory.

Consider first the case when the small mass in eq. (\ref{L_Psi})
is zero, $\mu = 0$. Then, the anomaly in the $(d-1)$ current
(\ref{Jmu}) tells us that the scattering process should produce
two massless fermions on the equator, with the total of 2 units of
chirality. On the other hand, the corresponding current of the
$d$-dimensional theory, eq. (\ref{JA}), is conserved exactly, so
there should be an additional contribution to the charge balance.

To see where this additional contribution comes from, recall that
in $d=3$ a vortex, in the presence of a single massive fermion
with mass $M$, acquires half a unit of the fermion number
\cite{Redlich:1983kn,Redlich:1983dv}:
\be
\langle J^A \rangle = \frac{M}{8\pi |M|} \epsilon^{ABC} F_{BC} \;
, \label{half-unit}
\ee
where $J^A$ is the current of that single species. This effect
occurs in the bulk of the $d=3$ spacetime, where the vortex is
initially positioned, and can be regarded as a result of the
polarization of the massive Dirac sea by the field $F_{BC}$.

In our Lagrangian (\ref{L_Psi}), there are two species of
fermions, with opposite signs of the mass. As a result, the gauge
charge of the vortex is now zero (so that in contrast to the
Callan-Harvey mechanism \cite{Callan:1984sa}, there is no net
Chern-Simons action), but the global charge, corresponding to the
current (\ref{JA}), is doubled. In addition, in the presence of a
domain wall, the mass $M$ for each fermionic species has opposite
signs in the two hemispheres. Thus, the global charge of the
vortex is now equal to $\pm 1$, depending on the hemisphere. So,
as the vortex crosses the equator, it produces two units of
chirality in the form of fermions bound to the wall, but its own
charge also changes, precisely by the opposite amount. In this
way, the exact conservation of the $d$-dimensional current is
reconciled with the anomalous production of fermions on the
equator.

Next, let us restore the small mass in eq. (\ref{L_Psi}), $\mu
\neq 0$. In this case, the conservation of the current (\ref{JA})
is no longer exact. Instead, we have
\be
\partial_A J^A_G = 2i\mu^* \bar{\Psi}_2  \Psi_1 - 2i \mu \bar{\Psi}_1 \Psi_2 \; .
\label{mass_terms}
\ee
The vortex states are no longer exact eigenstates of the
corresponding global charge,
\be
Q_G = \int J^0_G \sin\theta d\theta d\phi \; ,
\ee
but we can still consider averages of $Q_G$ in these states.
Specifically, let us consider the adiabatic limit, when the vortex
crosses the equator very slowly---the timescale of its motion is
much larger than $\mu^{-1}$. In this limit, fermions remain, to 
a good accuracy, in adiabatic vacuum.
Averaging eq. (\ref{mass_terms}) over this state and integrating
over the sphere and over an interval of time, we find
\be
\langle Q_G(t_2) \rangle - \langle Q_G(t_1) \rangle = \int
[2i\mu^* \langle \bar{\Psi}_2  \Psi_1\rangle - 2i \mu \langle
\bar{\Psi}_1 \Psi_2\rangle] \sin\theta d\theta d\phi dt \; .
\ee
The averages on the right-hand side can be obtained through the
derivatives of the anomalous action (\ref{anom_sphere}) with
respect to $\mu$ and $\mu^*$. In this way, we find
\be
\langle Q_G(t_2) \rangle - \langle Q_G(t_1) \rangle
 \approx \frac{1}{\pi} \int d\phi dt F_{0\phi} (0, \phi, t) \; .
\label{change_in_Q}
\ee
The approximation sign reminds us that in the action
(\ref{anom_sphere}) we have neglected terms of higher orders in
$\mu$. To the same accuracy, the average charges of the vortex
before and  after the equator crossing are still determined by eq.
(\ref{half-unit}): $\langle Q_G(t_1) \rangle\approx -1$, $\langle
Q_G(t_2) \rangle\approx 1$. We see that the change in $\langle
Q_G\rangle$, due to restructuring of the Dirac sea of the massive
modes, is precisely as required for eq. (\ref{change_in_Q}) to
hold. In fact, the flow of information could have been reverted:
we could have used the simple counting of charges to restore the
coefficient in the effective action (\ref{anom_sphere}).

The fact that for $\mu\neq 0$ the current $J^A_G~~$ is not exactly
conserved, and therefore vortex states are not eigenstates of the
charge, has important consequences for the realization of the
global $U(1)$ symmetry. In (2+1), this $U(1)$ is anomaly-free, and
we can use it to rotate the phase $\theta_M =\arg \mu$ to zero.
However, in quantum theory, this will also transform the state
vector. If we think of the vortex-antivortex state as a
superposition of components with different values of $Q_G$, the
transformation will change the relative phases of the components.
These relative phases then become a counterpart, in the (2+1)
theory with an non-anomalous $J^A_G$, to the vacuum $\theta$-angle
in the (1+1) theory with an anomalous $J^\mu_5$.

To summarize, for the case of a domain wall on a sphere, the
motion of vortices provides a very visual way to understand the
effect of spectral asymmetry. We have not developed a corresponding
visual tool for the case of an orbifold, but we expect that in that case 
the mismatch between the ``ultraviolet'' and ``infrared'' anomalies 
can similarly be attributed to restructuring of the massive part of 
the fermionic spectrum in a $z$-dependent field $F_{\mu\nu}$.

\section{Gauge dynamics and theta-vacua}
\label{sect:gauge_field} The presence, for $\mu  \neq 0$, of
instanton processes that do not produce any fermions implies a
possibility to have an observable $\theta$ angle. (This is similar
to how in QCD, to have a $\theta$-vacuum, all quarks should be
massive.) However, the presence of such processes is only one
necessary condition for the existence of $\theta$. The other
condition is that these processes connect states that are
degenerate, or nearly degenerate, in energy. If instantons have to
climb a high potential ladder, they will be blocked at low
energies.

Note that a description of gauge dynamics requires that we
construct scenarios where the gauge field splits into a low-energy
mode, corresponding to a  $(d-1)$-dimensional gauge field, and
high-energy modes separated from the low-energy mode by a large
gap. This is relatively straightforward to achieve on an orbifold,
and somewhat less straightforward on a sphere.  We now consider
these cases in turn.

\subsection{Theta-vacua on an orbifold}
\label{subsect:theta_on_orbifold} For our toy model in $2+1$
dimensions we take an Abelian gauge field, as it is for this
theory that the vacuum in $1+1$ dimensions has complicated
structure. We consider a massless gauge field with the Lagrangian
\be
L=-\frac{1}{4} F_{AB}F^{AB}~, \label{acel}
\ee
where $F_{AB} = \partial_A A_B - \partial_B A_A$. This massless
case is somewhat degenerate, since it has no physical propagating
mode in $1+1$ dimensions: the only physical mode in a massless
$1+1$ gauge theory is a uniform electric field and its canonically
conjugate coordinate given by the Wilson line  $\int dx A_1(x,z)$.
However, it is precisely the dynamics of this mode that is of
interest to us here. Indeed, a constant (in time) uniform electric
field plays the role of a $\theta$-angle in (1+1)
\cite{Coleman:1976uz}.

We consider this theory on a orbifold $O\times S_1$, where $O$ is
an interval of length $R$, and $S_1$ is a large circle of length
$L$. First, we consider the free gauge theory, defined by the bilinear
Lagrangian (\ref{acel}), and then add interaction with fermions.

Variation of the action, besides the ordinary Maxwell equations
\be
\partial_A F^{AB}=0
\label{max}
\ee
valid in the bulk, gives now the extra boundary terms,
\be
\int d^2x\left(\delta A^\mu(R/2,x)F_{2\mu}(R/2,x)-\delta
A^\mu(-R/2,x)F_{2\mu}(-R/2,x)\right)=0~,
\ee
which lead to boundary conditions \cite{footnote_bc}
\be
F_{\mu 2}\vert_{\pm R/2}=0~. \label{bc_gauge}
\ee
assuming arbitrary variations of $A_\mu$ at the boundaries. Note
that since $A_0$ plays the role of a gauge function, gauge
transformations with arbitrary continuous gauge functions are
admitted (see below for more detail).

The general solution to the free Maxwell equations (\ref{max})
with boundary conditions (\ref{bc_gauge}) has the form:
\ba
A_\mu &=& \frac{\partial \alpha(x^A)}{\partial x^\mu}
+\sum_{n=0}^\infty A_\mu^n(x^\nu) \chi_n(z)~, \label{Amu} \\
A_z&=& \frac{\partial \alpha(x^A)}{\partial z}\; , \label{Az}
\ea
where $\alpha$ is an arbitrary function reflecting the gauge
freedom, fields $A_\mu^n(x^\nu)$ satisfy the vector field equation
\be
\partial^\mu F_{\mu\nu}^n+M_n^2 A_\nu^n = 0 \; ,
\ee
where $F_{\mu\nu}^n=\partial_\mu A_\nu^n -\partial_\nu A_\mu^n$,
the mass $M_n$ is given by  eq. (\ref{mass}), and $\chi_n$ are
defined in (\ref{psch}). This decomposition is valid for the
five-dimensional case as well. The low-energy theory is just the
electrodynamics in  $1+1$ or $3+1$ dimensions.

In 3d, the field strengths for the solution (\ref{Amu}), (\ref{Az}) 
are
\ba
\nonumber E_1&=&E_1^0-\sum_{n=1}\frac{\partial
a_n(x,t)}{\partial t}M_n\chi_n(z)\; ,\\
\nonumber E_2&=&\sum_{n=1}\frac{\partial^2
a_n(x,t)}{\partial t\partial x}\psi_n(z)\; ,\\
B &=&\sum_{n=1}\left(\frac{\partial^2 a_n(x,t)}{\partial
x^2}-a_n(x,t)M_n^2\right)\psi_n(z)\; ,
\ea
where $a_n(x,t)$ satisfy the Klein-Gordon equation
$(\partial_\mu\partial^\mu+M_n^2)a_n(x,t)=0$. Thus, $E_1$ includes
a constant electric field $E_1^0$ in $x$ direction, which is a
Lorentz scalar in $1+1$ dimensions. The fact that this constant
electric field is allowed by the 3d equations of motion and by the
boundary conditions is essential for discussion of $\theta$ vacua
on orbifold. Note that in $1+1$ dimensional electrodynamics the
constant electric field is also a solution of equations of motion
and plays the role of a $\theta$ angle \cite{Coleman:1976uz}.

We now want to address the following questions: (i) Does the
complete (2+1)-dimensional theory have  a complicated vacuum
structure characterized by a vacuum angle $\theta$? (ii) Does the
phase $\theta_M$ of the fermion mass $\mu$ contribute to an
observable $\theta$-angle? As we will see, the answers to both of
these questions are affirmative.

We begin with constructing classical vacua, i.e., states of
minimal classical energy. Let us choose the gauge $A_0=0$ and
consider time-independent gauge transformations. The group of all
such transformations  consists of functions $\alpha$ for which
$\exp[i\alpha(z,x)]$  is continuous on $O\times S_1$. In addition
to ``local" or ``small" gauge transformations, for which functions
$\alpha(z,x)$ themselves are continuous, this condition allows for functions
$\alpha(z,x)$ that have a jump $2\pi n$ with integer $n$ along a
line in the $(x,z)$ plane. Such a line can either form a closed
loop or connect the opposite points of the interval $O$. In the
first case, the loop is contractible, and the transformation can be
continuously deformed to a small gauge transformation, but the
second case is a non-contractible, ``large'' transformation. For
$n=1$, these large gauge transformations can be reduced to the
form
\be
\alpha(z,x) = \left\{ \begin{tabular}{ll}
$\frac{2\pi}{L} [x -x_0(z) ]$ , & ~~~~~$0 \leq x \leq x_0(z)$ , \\
$\frac{2\pi}{L} [x -x_0(z) - L]$ , & ~~~~~$x_0(z) <  x \leq L$ ,
\end{tabular} \right.
\label{large}
\ee
where the function $x_0(z)$ defines a line (without intersections)
on which $\alpha$ jumps by $2\pi$. So, the $N$-vacuum is the 
gauge-field configuration
\be
A^{(N)}_1= \frac{2 \pi N}{e L}, ~~A^{(N)}_2 =  -\frac{2\pi N}{e
L}\frac{\partial x_0}{\partial z} \; ,
\ee
which is characterized by an integer Chern-Simons number
\be
N_{CS} = \frac{e}{2\pi}\int dx A^{(N)}_1(x,z) = N \; .
\ee
Note that for the vacuum configurations $N_{CS}$ does not depend
on $z$ (for an arbitrary gauge background that is not so)  and
that it is invariant under small gauge transformations for any
$A_1(x,z)$. Under the large gauge transformation (\ref{large}), it
changes by one.

This construction of the classical $N$-vacua on the orbifold is
almost identical to the similar construction in (1+1) dimensions. As
we now go to quantum theory, we construct a $\theta$-vacuum as a
linear superposition of states built near these classical vacua
\cite{Callan:1976je,Jackiw:1976pf}. The functional integral for
vacuum-vacuum transitions can be written as
\be
\int {\cal D} A_B \exp\left(i \act_3 -i  \frac{e\theta_{\rm
vac}}{4\pi}\int d^2x \epsilon_{\mu\nu}F^{\mu\nu}(z,x)\right) \; ,
\label{eff_action}
\ee
where $\act_3$ is the complete three-dimensional action together with
the necessary gauge fixing and ghost terms. Note that
\be
\int d^2x\epsilon_{\mu\nu}F^{\mu\nu}(z,x)
\ee
is $z$-independent for the vacuum-vacuum transitions. Thus, the
vacuum of a $U(1)$ gauge theory on orbifold is characterized by an
angle $\theta$, exactly in the same way as in the effective
(1+1)-dimensional theory. Clearly, this is related to the topology
of the orbifold: the mapping $O\times S_1 \rightarrow U(1)$ is
non-trivial and characterized by an integer $Z$, just as the
mapping $S_1 \rightarrow U(1)$ is in the low-energy theory. A
similar argument applies to 5d non-Abelian theories, for which the
space $O\times S_3$ has a non-trivial mapping to the group SU(2).

To detect a $\theta$-angle, we need to have charged particles.
So, let us include interaction of the gauge field with two species of
fermions, such as those described in Sect. \ref{sect:determinant}.
The fermions can be integrated out, and for small and slowly varying
$F_{AB}$, their main contribution to the effective action is given
by eq. (\ref{anom_slow}). So, in this approximation, the effective action
still has the form (\ref{eff_action}) but with $\theta_{\rm vac}$ replaced
by
\be
\theta_{\rm tot} = \theta_{\rm vac}+\theta_M \; ,
\label{theta_tot}
\ee
where $\theta_M$ is the phase of the fermion mass $\mu$.

\subsection{Vacuum structure on a sphere}
We have seen that in this case instanton fluctuations, which
activate the $\theta_M$ dependence, correspond to scattering of
vortices on the domain wall. The essential difference with the
case of the orbifold is that such a fluctuation now does not
connect two vacuum states. Rather, it connects the vacuum to a
state with a vortex in the northern hemisphere and an antivortex
in southern (or vice versa). This is consistent with topological
considerations: the mapping from $S_2$ to $U(1)$ is trivial, so
there are no ``large'' gauge transformations and no degenerate
$N$-vacua.

The question we want to address in this subsection is if there can
nevertheless be an effective $\theta$-angle, due to existence of
vortex states that are {\em nearly} degenerate with the vacuum.
This question needs to be answered within a scenario where the
effective low-energy theory is that of a (1+1)-dimensional gauge
field, while all other gauge modes have a large gap.

To construct such a scenario, we consider a family of Abelian
theories with a coupling constant dependent on the spherical angle
$\theta$:
\be
L = -{1\over 4} \sqrt{g} {1\over h(\theta)} g^{AB} g^{CD} F_{AC}
F_{BD} = \frac{\sin\theta}{2h} ( F_{0\theta}^2 + {1\over
\sin^2\theta} F_{0\phi}^2 - {1\over \sin^2\theta} F_{\theta\phi}^2
) \; , \label{L_A_sphere}
\ee
where $g_{AB} = {\rm diag} (1,-1,-\sin^2\theta)$ is the metric.
The $\theta$-dependent coupling $h(\theta) > 0$ will be referred
to as the warp factor. Such space-dependent couplings arise
naturally in brane-world scenarios
\cite{Oda:2000zc,Dubovsky:2000av,Dvali:2000rx}. 

Magnetic field $b$
has been defined in eq. (\ref{b_field}): $b = F_{\theta\phi} /
\sin\theta$. The time derivative of this expression gives
Faraday's law on the sphere:
\be
\dot{b} = \frac{1}{\sin\theta} (\partial_\theta \dot{A}_\phi -
\partial_\phi \dot{A}_\theta) = \frac{1}{\sin\theta}
(\partial_\theta F_{0\phi} - \partial_\phi F_{0\theta} ) \; .
\label{F-law}
\ee
Eq. (\ref{F-law}) shows that the total magnetic flux through the
sphere is conserved:
\be
{\cal B} = \int d\phi \int_0^{\pi} b \sin \theta d\theta = \const
\label{flux_sphere}
\ee
In what follows, we restrict ourselves to the sector with zero
flux,
\be
{\cal B} = 0 \; , \label{zero_flux}
\ee
i.e., we assume that there is no monopole inside the sphere.

Equations of motion following from (\ref{L_A_sphere}) are
\ba
\partial_0 (\frac{\sin\theta}{h} F_{0\theta} )
- \partial_\phi (\frac{1}{ h\sin\theta} F_{\phi\theta} )
& = & 0 \; , \\
\partial_0 (\frac{1}{ h \sin\theta} F_{0\phi} )
- \partial_\theta (\frac{1}{ h \sin\theta} F_{\theta\phi} )
& = & 0 \; , \\
\partial_\theta (\frac{\sin\theta}{ h} F_{\theta0} )
+ \partial_\phi (\frac{1}{ h \sin\theta} F_{\phi 0} ) & = & 0 \; .
\ea
Consider first solutions for which all $F_{AB}$ are
time-independent. Then, the first two of the equations of motion
reduce to $\partial_\phi b = 0$ and $\partial_\theta(b / h) = 0$,
which are solved by
\be
b = c_1 h \; , \label{monopole}
\ee
where $c_1$ is an arbitrary space- and time-independent
coefficient. This is the monopole solution characterized, for $c_1
\neq 0$, by a non-zero total flux. We have projected it away by
imposing the zero-flux condition (\ref{zero_flux}).

Next, consider solutions for which all $F_{AB}$ depend on time as
$e^{-i\omega t}$ with $\omega \neq 0$. Then, the electric fields
are
\ba
F_{0\theta} & = & \frac{\partial_\phi b}{i\omega \sin\theta} \; , \\
F_{0\phi} & = & - \frac{h\sin\theta}{i\omega} \partial_\theta
(b/h) \; . \label{elec_sphere}
\ea
Substituting these expressions into eq. (\ref{F-law}) and
expanding in the eigenstates of the angular momentum, we obtain a
closed equation for component of $b$ with angular momentum $m$
($m=$integer):
\be
-  \frac{1}{\sin\theta} \partial_\theta [ h \sin\theta
\partial_\theta (b/h) ] + \frac{m^2}{\sin^2\theta} b = \omega^2 b
\; . \label{eq_modes_sphere}
\ee
Defining $B = b /h$ and $H = h\sin\theta$, we can rewrite this
equation as
\be
\partial_\theta ( H \partial_\theta B ) = - (\omega^2 - m^2 /\sin^2\theta) H B \; .
\ee
Setting $B = \chi /\sqrt{H}$ and $H=e^f$, we rewrite it further as
a Schr\"{o}dinger equation
\be
\chi'' - [ \half f'' + {1\over 4} (f')^2 ] \chi = - (\omega^2 -
m^2 /\sin^2\theta) \chi \; . \label{S-eq}
\ee
Primes denote derivatives with respect to $\theta$. The ground
state of this Schr\"{o}dinger problem is $\chi \propto \sqrt{H}$.
This coincides with the monopole solution (\ref{monopole}), which
we have projected out. We are interested in the lowest-energy mode
satisfying the condition (\ref{zero_flux}).

Let us consider the case when all modes are concentrated mostly in
small regions near the poles. A simple choice of the warp factor
that leads to such an arrangement is
\be
H = C \exp (-\half \kappa^2 \sin^2\theta) \; ,
\label{warpH}
\ee
where $C$ is a constant. Taking $\kappa \gg 1$ and considering
only a vicinity of the north pole, $\theta \ll 1$, we see that in
this case the potential in (\ref{S-eq}) is approximately that of a
harmonic oscillator:
\be
V(\theta) = \half f'' + {1\over 4} (f')^2 \approx {1\over 4}
\kappa^4\theta^2 - \half \kappa^2 \; .
\ee
Upon replacement $\theta \to \pi - \theta$, we obtain a
corresponding expression near the south pole. The ground states of
these oscillators comprise the low-energy subspace of our system.
Due to the (exponentially small) overlap at the equator, these
ground states form symmetric and antisymmetric linear
combinations. The symmetric combination, which is the true ground
state of the system, is once again the monopole solution
(\ref{monopole}). The antisymmetric combination is the state we
are interested in: it has zero total flux and, if suitably
populated, corresponds to a vortex at the north pole and an
antivortex at the south pole. The fields in this state oscillate
at exponentially small frequency
\be
\omega_1 \sim \kappa \exp( - \const \times \kappa^2 {\cal R}^2) \; ,
\label{omega_1}
\ee
where we have restored the radius ${\cal R}$ of the sphere.
All other modes are separated from this one by the gap $\omega_2^2
\approx \kappa^2$.

The above spectrum is reminiscent of the one that occurs in models
that use a warped gauge coupling as a means to obtain light vector
bosons \cite{Shaposhnikov:2001nz}. The crucial difference is that
in our case the exponentially light mode occurs only for angular
momentum $m=0$. So, it does not correspond to a vector particle
propagating along the domain wall. Rather, it is the counterpart
of the ``topological'' mode, for which $A_\phi$ and $F_{0\phi}$
are constant along ``our'' dimension. If that mode were constant
in time, it would correspond to a conventional $\theta$-angle,
just as in (1+1) dimensions or in the case of orbifold. We see,
however, that on the sphere this mode acquires a small but nonzero
frequency, resulting in a variation of the effective
$\theta$-angle with time.

In a static universe, the case for which the above results have
been obtained, the dynamics of the ``topological'' mode is
oscillatory. In an expanding universe, we expect this mode to be damped
by the expansion. Furthermore, if $\omega_1$ is not particularly small, 
and the gauge
field interacts with light matter, the oscillations of the
``topological'' mode can decay into matter particles. For QCD in
4d, either of these scenarios constitutes a solution to the
strong-CP problem.

If $\omega_1$ is, in fact, small (as in the above example, where it is
suppressed exponentially by the size of ``our'' dimensions), 
$\omega_1^{-1}$ may well be a cosmological timescale, so the relaxation
of the effective $\theta$-angle and of the associated vacuum energy
will occur relatively late in the cosmological history.

\subsection{Effective Lagrangian for a time-dependent $\theta$}
As follows from eq. (\ref{elec_sphere}), the ``topological'' mode, 
oscillating at frequency $\omega_1$, corresponds to an oscillating (in time) 
and uniform (in $\phi$) electric field on the equator. We know that in (1+1) 
dimensions or in the case of orbifold a constant
electric field is the classical counterpart of a $\theta$-angle 
\cite{Coleman:1976uz}. In quantum theory on the orbifold,
the $\theta$ dependence can be described by the effective Lagrangian
appearing in eq. (\ref{eff_action}). One may expect that 
a similar description, but with an effective, time-dependent 
$\theta$-angle, exists in the case of a sphere.

Such a description is provided by the following dimensionally reduced 
action for the fields on the equator:
\be
\act = \act_2 + \frac{1}{2\omega_1^2} \int
dt \left( \frac{da}{dt} \right)^2 + \frac{1}{\sqrt{2\pi}} \int dt d\phi 
a(t) F^{(2)}_{0\phi} \; ,
\label{action_sphere}
\ee
where $\act_2$ is the action for the theory on a circle, and $F^{(2)}_{0\phi}$
is the canonically normalized field strength of that theory. 
The quantum-mechanical variable $a(t)$ depends only on time but not on space:
it is a dual representation of the ``topological'' mode of
the gauge field; $\omega_1$ is the
frequency of that mode, eq. (\ref{omega_1}).

If the action $\act_2$ contains fermions, we integrate them out and, for small,
slowly-changing field strengths, obtain as the leading terms the anomalous action
(\ref{anom_sphere}),
proportional to the phase $\theta_M$ of the fermion mass. 
We see, however, that in the present case this anomalous action can be 
absorbed by a shift in the variable $a$. In other words, it only changes the
initial conditions for $a(t)$. The same applies to any term of the form
\be
\act'_3 = \int dt d\theta d\phi u(\theta) F_{0\phi} \; ,
\label{act'_3}
\ee
which we might have added by hand to the original 3d action 
($u$ is some function). This is because at low energies $F_{0\phi}$ projects 
onto the ``topological'' mode, so (\ref{act'_3}) becomes of the same form as the
last term in (\ref{action_sphere}).
So, in what follows we assume that $a$ in (\ref{action_sphere}) already
includes the effect of $\theta_M$ and of any term such as (\ref{act'_3}), 
i.e., we set $\theta_M = 0$ and $u=0$. (Strictly speaking, this requires 
that the combined
$a(t)$ is sufficiently small to prevent the decay of the uniform electric field,
eq. (\ref{ave_F}) below, into fermion pairs, cf. ref. \cite{Coleman:1976uz}.)

Now, integrating out the uniform component of $A^{(2)}_\phi$,
we obtain
\be
\langle F^{(2)}_{0\phi} \rangle = - \frac{1}{\sqrt{2\pi}} a(t) 
\label{ave_F}
\ee
and a simple oscillatory Lagrangian for $a(t)$:
\be
L_a = \frac{1}{2\omega_1^2} \left( \frac{da}{dt} \right)^2 - \half a^2 \; .
\label{L_a}
\ee
We see that in the limit $\omega_1 \to 0$, the inertia of $a$
grows indefinitely. Formally setting $\omega_1 = 0$ would convert
$a$ into a conventional time-independent $\theta$-angle.

The condition that $F^{(2)}_{0\phi}$ is slowly-varying implies that
$\omega_1 \ll |\mu|$, where $\mu$ is the fermion mass. 
For large enough $\omega_1$, the oscillating $a$ can efficiently decay
into fermions, and eq. (\ref{L_a}) is no longer applicable. 
(A single quantum of $a$ can decay into
fermions when $\omega_1 > 2|\mu|$, two such quanta will be required
when $|\mu| < \omega_1 \leq 2|\mu|$, etc.)

The variable $a(t)$ can be viewed as a ``global axion'', in the
sense that it couples to the topological density in a way similar
to how the usual axion \cite{Peccei:1977hh,Peccei:1977ur} does. 
However, since $a$ only depends on
time, and not on space, it does not correspond to a new particle.
In fact, it is not even an additional degree of freedom, external
to the original 3d theory: as eq. (\ref{ave_F}) shows, it is simply a different
representation of the time-dependent uniform electric field.

Clearly, existence of such a variable would not be possible in a
perfectly Lorentz-invariant theory but, of course, the 2d Lorentz
invariance is not exact in our brane-world scenario.

Finally, we note that although eq. (\ref{ave_F}) is specific to a
3d $U(1)$ gauge theory, the general structure of eq.
(\ref{action_sphere}) is not. For a 5d non-Abelian theory (with
$\sp_{d-1} = S_4$), we would replace (\ref{action_sphere}) with
\be
\int {\cal D} A_B \exp\left\{ i \act_4 + i \frac{1}{2\omega_1^2} \int
dt \left( \frac{da}{dt} \right)^2 + i v \int d^4 x a(t) 
F_{\mu\nu}\tilde{F}^{\mu\nu} \right\} \; , \label{nonab_sphere}
\ee
where $\act_4$ is the conventional 4d action, and $v$ is a suitably chosen 
constant.

\section{Theory on a disk}
\label{sect:disk} Our results on chiral fermions and the vacuum
structure on a sphere are related to the topology, rather than geometry,
of the spatial manifold. Similar results are valid for a simpler, flat
geometry. Simply cut a sphere along the equator, choose a
hemisphere, and make it flat by replacing it with a disk. Then,
substitute the domain wall by a suitable boundary condition. In
this subsection we present the corresponding equations. We will
call the boundary of the disk the brane and its interior the bulk.

\subsection{Fermions on a disk}

Introduce Cartesian coordinates $x$ and $y$ with origin at the
center of the disk. Three-dimensional $\gamma$-matrices used in
this subsection are $\gamma^0=\tau_3,~\gamma^1=i \tau_1,~ \gamma^2
= i \tau_2$. Note that these matrices are associated with the
Cartesian coordinates. Then, the Dirac equation $i
\gamma^\mu\partial_\mu \Psi -M\Psi=0$, where $M>0$ is a constant
fermion mass in the bulk, can be written in polar coordinates ($ x
= r \cos\phi,~y=r\sin\phi$) in the form of a Schr\"{o}dinger
equation $i \frac{\partial \Psi}{\partial t} = H\Psi$ with the
Hamiltonian
\be
H = \left( \begin{tabular}{lr}
$M$&$e^{-i\phi}\left(-\partial_r+\frac{i}{r}\partial_\phi\right)$ \\
$e^{i\phi}\left(\partial_r+\frac{i}{r}\partial_\phi\right)$&$-M$
\end{tabular} \right) \; ,
\ee
leading to the energy eigenvalue problem $H \Psi =E \Psi$. The
regular at $r=0$ solutions are:
\be
\left(\begin{tabular}{l}
$\psi_1$\\
$\psi_2$
\end{tabular} \right)=
\left(\begin{tabular}{l}
$~~~~~~~~~e^{in\phi}J_n(k r)E_+$\\
$-e^{i(n+1)\phi}J_{n+1}(k r) E_-$
\end{tabular} \right)
 \; ,
\ee
for $E^2>M^2$ and
\be
\left(\begin{tabular}{l}
$\psi_1$\\
$\psi_2$
\end{tabular} \right)=
\left(\begin{tabular}{l}
$~~~~~~~e^{in\phi}I_n(k r)E_+$\\
$e^{i(n+1)\phi}I_{n+1}(k r) E_-$
\end{tabular} \right)
 \; ,
\ee
for $E^2<M^2$. Here $E_+=\sqrt{|E+M|},~~E_-=\sqrt{|E-M|},~~k
=\sqrt{|E^2-M^2|}$, $J_n$ and $I_n$ are the Bessel and modified
Bessel functions, $n$ is an integer.

A boundary condition that produces a left chiral fermion on the
brane and is consistent with the hermiticity of the Hamiltonian is
\be
(1-\gamma_5(\phi))\Psi\vert_{r=R}=0, ~~\mbox{or}~~ \psi_2=
e^{i\phi}\psi_1~, \label{lef}
\ee
where $\gamma_5(\phi) = -i (\gamma^1 \cos\phi + \gamma^2
\sin\phi)$ is a chirality matrix in polar coordinates. Eq.
(\ref{lef}) leads to the eigenvalue equations
\ba
\label{first}
I_n(k R) E_+&=&I_{n+1}(k R) E_-,~~ E^2<M^2~,\\
J_n(k R) E_+&=&-J_{n+1}(k R) E_-,~~ E^2>M^2~, \label{second}
\ea
where $R$ is the radius of the disk. Solution to eq.
(\ref{first}), at $M R \gg 1$ gives a chiral mode with dispersion
relation $E \approx -(n+\frac{1}{2})/R$, exactly as we have
obtained for a sphere. This mode is localized at the boundary of
the disk, with an exponential wave function $\sim e^{-M(R-r)}$ for
$(R-r)/R \ll 1$. Solutions to eq. (\ref{second}) lead to massive
bulk modes with energies greater than $M$.

A right-handed fermion can be derived in a similar manner, by
choosing the negative mass parameter $M<0$ and by changing the
boundary condition (\ref{lef}) to
$(1+\gamma_5(\phi))\Psi\vert_{r=R}=0$.

A massive (with mass $\mu$), Dirac fermion living on the
boundary of the disk can be introduced exactly in the way it has
been done for the orbifold or a sphere, namely by including two
fermions, the first one ($\Psi_1$) producing the left one and the
second ($\Psi_2$) producing the right fermion, with the mixing
mass term $\mu\bar{\Psi}_1\Psi_2$.

\subsection{Gauge fields on a disk}
Similarly to the case of a sphere, a gauge field that has a
(1+1)-like low-energy mode, while other modes are separated by a
large gap, can be introduced through Lagrangian
\be
L = -\frac{1}{4}\Delta(r) F_{AB}F^{AB}~, \label{warp_disk}
\ee
where the warp factor $\Delta(r)$ is of order one in a small
vicinity of the disk boundary and goes to zero at $r \rightarrow
0$. A typical model for $\Delta(r)$ could be
\be
\Delta(r) = \left(\frac{r}{R}\right)^2 e^{-M(R-r)}~.
\ee
As for the orbifold case, the boundary condition to the gauge
field is
\be
F_{r\phi}\vert_{r=R} = F_{r0}\vert_{r=R}=0~.
\ee
However, in contrast to the orbifold case, the vacuum is
topologically trivial, as follows from the fact that the boundary
of the disk is a simply connected manifold. Namely, allowed gauge
functions may contain a $2 \pi n$ jump along a closed loop on the
disk or along a line connecting two points at the boundary. All
these transformations can be continuously transformed into trivial
gauge transformations.

The absence of a conventional $\theta$-angle on a disk still
leaves us with the possibility to have an effective
$\theta$-angle, due to transitions that connect the vacuum to a
vortex state. The only difference with the sphere in this regard
is that the total magnetic flux through the disk is not conserved.
But this is in fact necessary for a candidate vortex state to
be connected to the vacuum: the disk is analogous to a hemisphere, 
rather than the entire sphere in our previous example. 
For a suitable warp factor in (\ref{warp_disk}), the
lowest-energy vortex state can be light enough to produce a
slowly-changing $\theta_{\rm eff}(t)$.

\subsection{Scalar fields on a disk}

Scalar fields can be localized on the boundary of a disk similarly
to fermions. In this subsection, we consider a real scalar field
as a prototype for a (complex) field that could give rise to
the Higgs mechanism.

We start from the standard Lagrangian
\be
L = \frac{1}{2}(\partial_\mu \varphi)^2 -\frac{M^2}{2} \varphi^2.
\ee
The regular at $r=0$ (in polar coordinates) solutions to the
equations of motion  are
\be
\varphi = e^{-i E t -i n \phi} \left\{
\begin{tabular}{ll}
$I_n( k r) \; ,$ & ~~~$E^2 < M^2$~, \\
$J_n( k r) \; ,$ & ~~~$E^2 > M^2$~.
\end{tabular} \right.
\ee
The boundary condition
\be
\left(\frac{\partial \varphi}{\partial r} -
\sqrt{M^2-m^2}\varphi\right)\vert_R=0~,
\ee
where $m^2 \ll M^2$, leads to the following dispersion relations:
\be
\frac{1}{2}(I_{n+1}(k R)+I_{n-1}(k R))=
\sqrt{\frac{M^2-m^2}{M^2-E^2}}I_{n}(k R),~~~~~E^2 < M^2 \; ,
\ee
\be
\frac{1}{2}(J_{n-1}(k R)-J_{n+1}(k R))=
\sqrt{\frac{M^2-m^2}{E^2-M^2}}J_{n}(k R),~~~~~E^2 > M^2 \; ,
\ee
which single out a light mode with dispersion relation 
(in the physically interesting limit $R\rightarrow \infty,
~n\rightarrow \infty,~n/R = \const$, with $m$ and $M$ fixed)
\be
E^2 \approx \left(\frac{n}{R}\right)^2 + m^2
\ee
and a wave-function localized on the brane. Other, bulk modes, have
large masses and are non-observable at small energies.

\section{Discussion and Conclusions}
\label{sect:conclusions} Results of this paper are two-fold.
First, we have shown how conservation of a global current in odd
dimensionalities can be reconciled with the presence of an anomaly
in the reduced, even-dimensional theory. The central observation
here is the presence of an additional contribution to the charge
balance, due to restructuring of the massive fermion modes.

Second, we have presented a brane-world scenario that leads to a
time-dependent effective $\theta$-angle. In this scenario, the
space is a sphere, and a domain wall is positioned along the
equator. Since the mapping from the sphere to the gauge group is
trivial, the usual, time-independent $\theta$-angle is absent.
However, the requirement that the low-energy limit is a
dimensionally reduced gauge theory automatically brings in an
effective $\theta$-angle.

We have discussed this scenario in detail for the case of a $U(1)$
gauge field in a $(2+1) \rightarrow (1+1)$ compactification, see
Fig. \ref{fig:sphere}. In this case, a simple picture of the
effective $\theta$-angle can be obtained, based on tracking the
motion of flux between the two hemispheres. For a free gauge field
(but with a warped action), when
exact results can be obtained, the dynamics of $\theta_{\rm eff}$
turns out to be oscillatory. We expect that when the gauge field 
interacts with light matter or in a non-static universe these oscillations
will be damped, so that $\theta_{\rm eff}$ relaxes to zero. We
also expect that a similar relaxation dynamics will obtain for the
$(4+1) \rightarrow (3+1)$ compactification of a non-Abelian
theory, thus providing a solution to the strong-CP problem.

The idea that the vacuum structure of a gauge theory can be
modified in the presence of extra dimensions is by itself not new.
Indeed, already quite a while ago \cite{Khlebnikov:1987zg} we
pointed out that if the higher-dimensional theory is defined on a
space manifold $\sp_{d-1}$, which is compact and obeys the
property $\pi_3(\sp_{d-1})=1$, the vacuum is topologically trivial
and that this can be a basis for a solution to the strong-CP
problem.

Let us compare the structure of the manifolds of ref.
\cite{Khlebnikov:1987zg} and of the present work. We start from
the $2+1 \rightarrow 1+1$ compactification. In both cases the
topology of the space is that of a 2-sphere. In ref.
\cite{Khlebnikov:1987zg}, we proposed that the low-energy theory
is $1+1$ dimensional because the manifold has the form of a
sausage, with $L \gg R$, see Fig. \ref{fig:sausage}.

In this setup, there is no complete translational invariance along
``our" dimensions because of the presence of two highly curved
regions, where our space ``ends". Nevertheless, if an observer
resides far from these regions, the low-energy physics looks $1+1$
dimensional, as the size of extra dimension $R$ is assumed to be
small. This setup solves the $\theta$ problem in the following
way. First, the topology of space is such that no non-trivial
gauge transformations exist. Second, the determinant of the
fermionic mass matrix is always real, because the fermions are
vectorlike. The generalization of this picture to $4+1$ dimensions
has qualitatively the same features.

Note that compactness of the higher-dimensional space is
essential: only in this case the topological argument is
unambiguous. (Thus, for instance, a recent proposal
\cite{Chaichian:2001nx} for solving the strong-CP problem with a
non-compact manifold will not work.) The easiest way to see the
role of compactness is to step back to the 3d~$\rightarrow$~2d
case. In this case, the $\theta$-angle corresponds to a
time-independent electric field \cite{Coleman:1976uz}. On a
non-compact manifold, there is always a choice of boundary
conditions at infinity for which such a time-independent solution
can be found. As long as no a priori way to reject these boundary
conditions is proposed, the $\theta$ problem is not solved.

Disadvantages of a sausage-like manifold are quite obvious: it
breaks the translational invariance in a very peculiar way, and it
is far from being obvious that a structure like this can arise as
a solution of the Einstein equations when gravity is incorporated.
Moreover, one cannot include chiral fermions, and therefore
possibility of construction of a phenomenologically acceptable
electroweak theory is doubtful.

In the present paper, we have proposed another structure, which
solves the above-mentioned problems. First, the manifold of the type
shown in Fig. \ref{fig:sphere}, where the standard-model fields
are localized on a brane, leads to physics that is translationally
invariant in ``our'' dimensions (i.e., along the equator). There
is a trivial breaking of the Lorentz invariance, since our space
is compact, but this is suppressed by the size of our dimensions.
Moreover, a similar geometry can be obtained as a solution to the
Einstein equations, as was demonstrated in ref.
\cite{Gruppuso:2004db}. In that solution, two slices of AdS space
are glued together along a three-dimensional sphere representing
the observable space. Finally, the presence of a domain wall leads
naturally to chiral fermions and thus to possibility to construct
a realistic theory.

A convenient way to visualize the dynamics of the low-energy mode
that plays the role of an effective $\theta$-angle in this setup
is through its dual---the ``global'' axion introduced in eqs.
(\ref{action_sphere}) and (\ref{nonab_sphere}). This ``global
axion'' is very different form the usual axion in that it does not
depend on space and therefore does not correspond to a new
particle. Such a global axion is not subject to any astrophysical
constraints, as it cannot be excited in stars, whereas the
cosmological constrains for it may remain in force.

The timescale of changes in $\theta_{\rm eff}$ is controlled by the
size of extra dimensions and can easily be very much larger than
the inverse of the QCD mass scale $\Lambda_{\rm QCD}$. In this case,
all the standard QCD dynamics---except for the strong-CP 
problem---remains intact. In particular, the mechanism that gives mass
to the $\eta'$ meson is unaffected by the presence of the global axion,
regardless of whether one associated this mechanism with instantons or
any other non-perturbative fluctuations in the QCD vacuum.

Such a global axion may look bizarre from the point of view of
relativistic field theory, but as we have shown in this paper it
may be quite natural in higher-dimensional theories. Thus, the
absence of strong CP violation may indicate that the number of
spatial dimensions in our world is greater than three, and
moreover that the space has certain topological properties and is
compact.

We thank A. Boyarsky, T. Clark, S. Dubovsky, E. Roessl and  O.
Ruchayskiy for interesting discussions. S.K. thanks EPFL, where
part of this work was done, for hospitality. The work of S.K. was
supported in part by the U.S. Department of Energy through Grant
DE-FG02-91ER40681 (Task B). The work of M.S. was supported in part
by the Swiss Science Foundation.

\newpage
\appendix
\section{Green functions on the orbifold}

The computation of anomalies requires computation of several
Feynman diagrams. In this appendix we construct the relevant
fermionic propagators for the theory defined by Lagrangian
(\ref{L_Psi}).

Let us define for this end two scalar propagators,
$G_D(x^\mu;z,z')$ and $G_N(x^\mu;z,z')$ which satisfy the
equations
\ba
\nonumber \left[\partial_\nu\partial^\nu -\frac{d^2}{dz^2} +
m^2(z)+\frac{dm}{dz}+|\mu|^2\right]G_D(x^\mu;z,z')&=&\delta^2(x)\delta(z-z')~,\\
\left[\partial_\nu\partial^\nu -\frac{d^2}{dz^2} +
m^2(z)-\frac{dm}{dz}+|\mu|^2\right]G_N(x^\mu;z,z')&=&\delta^2(x)\delta(z-z')
\label{green}
\ea
and boundary conditions
\be
G_D(x^\mu;\pm
R/2,z')=0,~~\left(\frac{d}{dz}+m(z)\right)G_N(x^\mu;z,z')|_{\pm
R/2}=0~. \label{bcG}
\ee
They can be expressed through the orthogonal sets of functions
$\psi_n$ and $\chi_n$ defined in (\ref{eig}) as follows:
\ba
\nonumber G_D(x^\mu;z,z')&=&\int\frac{d^2
k}{(2\pi)^2}e^{ikx}\tilde G_D(k;z,z'),~~ \tilde
G_D(k;z,z')=\sum_{m=1}^\infty
\frac{\psi_m(z)\psi_m(z')}{-k^2+M_n^2+|\mu|^2}~,\\
G_N(x^\mu;z,z')&=&\int\frac{d^2 k}{(2\pi)^2}e^{ikx}\tilde
G_N(k;z,z'),~~ \tilde G_N(k;z,z')=\sum_{m=0}^\infty
\frac{\chi_m(z)\chi_m(z')}{-k^2+M_n^2+|\mu|^2}~. \label{gdgn}
\ea
The Green functions in Fourier space $G_D(k;z,z')$ and
$G_N(k;z,z')$ satisfy the equation
\be
\left[-\frac{d^2}{dz^2} +
m^2(z)+\frac{dm}{dz}+|\mu|^2-k^2\right]G_{D,N}(x^\mu;z,z')=\delta(z-z')~
\label{Fou}
\ee
and boundary conditions following from (\ref{bcG}).

A helpful relation between the two functions is
\be
(\partial_z-m) G_D(x^\mu;z,z')= (-\partial'_z-m) G_N(x^\mu;z,z')~.
\label{help}
\ee
In five-dimensional theory one simply replaces $\frac{d^2
k}{(2\pi)^2}\rightarrow \frac{d^4k}{(2\pi)^4}$. The integral over
spatial components of momentum should be understood as a sum over
Fourier harmonics since we work on a compact torus.

The free fermion propagator
\be
S=\left( \begin{tabular}{lr}
$S_{11}$ & $S_{12}$ \\
$S_{21}$ & $S_{22}$
\end{tabular} \right) \; ,
\ee
obeys the equation
\be
\left( \begin{tabular}{lr}
$i \gamma^A \partial_A  + m(z)$ & $-\mu$\\
$-\mu^*$ & $i \gamma^A \partial_A  - m(z)$
\end{tabular} \right)S = \delta^2(x)\delta(z-z')
\ee
and the boundary conditions
\be
P_L S_{1i}\vert_{z=\pm R/2}=0~,~~ P_R S_{2i}\vert_{z=\pm
R/2}=0~,~~i=1,2~~.
\ee
It is easy to check that it can be expressed through scalar
propagators $G_D$ and $G_N$ as
\ba
\nonumber S_{11}&=&-(i \gamma^A \partial_A  - m(z))\left[P_L
G_D+P_R G_N\right],~~
S_{21}=\mu^*\left[P_L G_D+P_R G_N\right]~,\\
S_{22}&=&-(i \gamma^A \partial_A  + m(z))\left[P_L G_N+P_R
G_D\right],~~ S_{12}=\mu~\left[P_L G_N+P_R G_D\right]~,
\label{propag}
\ea
where $P_L=\frac{1}{2}(1+\gamma_5)$ and
$P_R=\frac{1}{2}(1-\gamma_5)$ are the chirality projectors.

Now we construct explicitly the Green function  $\tilde
G_D(k;z,z')\vert_{k=0,\mu=0}\equiv G_D(z,z')$ needed for a number
of applications. It satisfies the equation
\be
\left[-\frac{d^2}{dz^2} + m^2(z)+\frac{dm}{dz}\right]G_D(z,z')=
\delta(z-z') \label{GD}
\ee
and the boundary conditions $G_D(\pm R/2,z')=0$.

Let us define the function
\be
\rho(a,b)=\int_a^b \chi_0(z)^2~dz~,
\label{defrho}
\ee
where $\chi_0(z)$ is the zero mode defined in (\ref{zero}). The
obvious properties of the function $\rho(a,b)$ are:
$\rho(-R/2,R/2)=1,~\rho(a,a)=0$. Then one can easily check that
the function
\be
G_D(z,z')=\frac{1}{2\chi_0(z)\chi_0(z')}
\left[\rho(-R/2,z)\rho(z',R/2)\theta(z'-z)+
\rho(-R/2,z')\rho(z,R/2)\theta(z-z')\right]
\ee
satisfies eq. (\ref{GD}) and the boundary conditions.

Another function we will need is
\be
G_D(z)\equiv
G_D(z,z)=\frac{\rho(-R/2,z)\rho(z,R/2)}{\chi_0(z)^2}~.
\ee

Now, we give the explicit expressions for the Green functions
$\tilde G_D$ and $\tilde G_N$ in a theory with $m(z)=0$:
\ba
\nonumber \tilde G_D(0,z,z')=\frac{1}{\mu \sinh\mu R}
\left[\sinh\mu(z+R/2)\sinh\mu(z'-R/2)\theta(z'-z)+\right.&&\\
\left. \sinh\mu(z-R/2)\sinh\mu(z'+R/2)\theta(z-z')\right]&&
\ea
\ba
\nonumber \tilde G_N(0,z,z')=-\frac{1}{\mu \sinh\mu R}
\left[\cosh\mu(z+R/2)\cosh\mu(z'-R/2)\theta(z'-z)+\right.&&\\
\left.\cosh\mu(z-R/2)\cosh\mu(z'+R/2)\theta(z-z')\right]&&~.
\ea

Finally, we construct the Green function $G_D(k;z,z')$ for a sharp
domain wall residing in an infinite space-time. For this we put in
(\ref{Fou})
\be
m(z) = m_0 \epsilon(z)
\ee
and define a function
\be
\Phi(z)= \theta(z)e^{-E(k) z} + \theta(-z)\left(e^{-E(k)
z}-\frac{m_0}{m_0+E(k)}e^{E(k) z}\right)~,
\ee
where
\be
E(k)=\sqrt{m_0^2+\mu^2-k^2}~.
\ee
Then
\be
G_D(k;z,z')=\theta(z-z')\Phi(z)\Phi(-z')+\theta(z'-z)\Phi(z')\Phi(-z)~.
\ee

\section{Derivation of eq. (\ref{cur})}
In this appendix we will show that the flux of the gauge current
depends on the value of the gauge field at the end points of
interval only. Note that in 3d no regularization is needed as all
integrals are convergent.

For slowly varying in $x^\mu$ gauge fields the expression for the
current can be written as
\ba
\label{cursl} J^2(x^\mu,z)= \frac{e}{2} \int d^2x' dz'
\epsilon_{\mu\nu}F^{\mu\nu}(x',z')\times\\
\nonumber x'^\alpha\left[
 \partial_\alpha G_D(x';z,z')(\partial'_z+m)G_N(x';z',z)\right.-&&\\
\nonumber \left.
\partial_\alpha G_N(x';z,z')(\partial'_z-m)G_D(x';z',z)\right]~&&.
\ea
The integral over $x'$ can be performed (going first to momentum
space) to give
\be
J^2(x^\mu,z)= \frac{e}{2\pi} \int
dz'\sum_{m=0}\sum_{n=1}G_{mn}(z,z') M_n
F(M_m,M_n)\epsilon_{\mu\nu}F^{\mu\nu}(x,z')~, \label{smc}
\ee
where
\be
G_{mn}(z,z')=\left[\chi_m(z)\chi_m(z')\chi_n(z')\psi_n(z)+
\psi_m(z)\psi_m(z')\psi_n(z')\chi_n(z)\right]
\ee
and
\be
F(M_m,M_n)=\frac{1}{(M_m^2-M_n^2)^2}
\left[M_n^2-M_m^2+M_m^2\log\left(\frac{M_m^2}{M_n^2}\right)\right]~.
\ee

Now, if one puts $z=\pm R/2$ directly in (\ref{cursl}) one gets
zero because the expression for the current contains $\psi_n(z)$
which is zero at the end points because of the boundary
conditions. Nevertheless, the limit $\lim_{\epsilon\rightarrow
+0}J^2(\pm(R/2-\epsilon))$ is not equal to zero. The reason is
that in spite of the fact that the sum in  (\ref{smc}) converges
for any $z,z'$, its derivative with respect to $z$ does not
converge at $z=\pm R/2$. As this is an ultraviolet effect which
involves infinite sums, the limit of small $\epsilon$ can be
computed in a theory without mass term $m(z)$: the wave function
$\psi_n$ and $\chi_n$ approach their limit (\ref{psch}) for large
$n$.

With this in mind we have:
\ba
&&\Delta J\equiv J^2(x^\mu,R/2-\epsilon)-J^2(x^\mu,-R/2+\epsilon)=
\frac{e}{2\pi} \int dz'\epsilon_{\mu\nu}F^{\mu\nu}(x,z')\frac{4}{R^2}\times\\
&&\sum_{m+n=even}\sin\frac{\pi n \epsilon}{R} \left[\cos \pi
(n-m)(\frac{z'}{R} -\frac{1}{2}) F_s(M_m,M_n)+ \cos \pi
(n+m)(\frac{z'}{R} -\frac{1}{2}) F_a(M_m,M_n)\right]~, \nonumber
\ea
where
\be
F_{s,a}(M_m,M_n)= \frac{1}{M_n\pm M_m}\left[1 \pm \frac{M_m
M_n}{M_n^2-M_m^2}\log\frac{M_n^2}{M_m^2}\right]
\ee
are symmetric (s) and antisymmetric (a) functions respectively.

The gauge field $A_\mu$ can be expanded over a complete set of
orthogonal functions on the interval as follows:
\be
\frac{e}{2\pi}\epsilon_{\mu\nu}F^{\mu\nu}(x,z')= \sum_{k=0}B_k(x)
\cos 2\pi k(\frac{z'}{R} -\frac{1}{2})+ \sum_{k=1}C_k(x) \sin 2\pi
k(\frac{z'}{R} -\frac{1}{2})
\ee
so that
\be
\Delta J=\frac{2}{R} \sum_{k=0}B_k(x) \left[ \sum_{n=2k}^\infty
\sin\frac{\pi n \epsilon}{R} F_s(M_{n-2k},M_n) +\sum_{n=0}^{2k}
\sin\frac{\pi n \epsilon}{R} F_a(M_{2k-n},M_n)\right]~.
\ee
For any fixed finite $k$ the limit of the second term is equal to
zero as the sum over $n$ contains a finite number of terms. On the
contrary,
\be
\lim_{\epsilon\rightarrow 0} \sum_{n=2k}^\infty \sin\frac{\pi n
\epsilon}{R} F_s(M_{n-2k},M_n)= \lim_{\epsilon\rightarrow 0}
\sum_{n=2k}^\infty\sin\frac{\pi n
\epsilon}{R}\frac{1}{M_n}=\frac{R}{2}
\ee
since the sum over $n$ can be replaced by an integral for small
$\epsilon$. Finally,
\be
\Delta J= \sum_{k=0}^\infty B_k(x)=
\frac{e}{4\pi}\epsilon_{\mu\nu}
\left[F^{\mu\nu}(x,R/2)+F^{\mu\nu}(x,-R/2)\right]~.
\ee

\end{document}